\documentclass[a4paper,11pt]{article}
\pdfoutput=1 

\usepackage{jheppub, amsmath, amssymb,amsthm,dsfont,graphicx,mathtools,multirow,rotating,placeins,verbatim} 
\usepackage[T1]{fontenc} 

\usepackage{tensor}
\usepackage[T1]{fontenc} 
\usepackage[all,color]{xy}
\usepackage{float}

\newcommand{\black}{\color{black}}

\newcommand{\Av}{\mathcal{A}}

\def\be{\begin{equation}}
\def\ee{\end{equation}}
\def\ba{\begin{eqnarray}}
\def\ea{\end{eqnarray}}
\def\bi{\begin{itemize}}
\def\ei{\end{itemize}}
\newcommand{\beq}{\begin{eqnarray}}
\newcommand{\eeq}{\end{eqnarray}}

\def\A{\mathcal{A}}
\def\zb{\bar{z}}

\def\Zb{\bar{Z}}
\def\L{\mathcal{L}}

\def\I{\mathcal{I}}

\newcommand{\Op}{\mathcal{O}}
\newcommand{\Li}{\mathcal{L}}
\newcommand{\pb}{\mathfrak{a}\mathfrak{d}}
\newcommand{\Ib}{\mathcal{I}}

\title{\boldmath Radiative phase space extensions at all orders in $r$ for self-dual Yang-Mills and Gravity }

\author[a,1]{Silvia Nagy,\note{Corresponding author.}}
\author[b]{Javier Peraza}

\affiliation[a]{Centre for Particle Theory, Department of Mathematical Sciences,\\
Durham University, Durham, DH1 3LE, U.K.}
\affiliation[b]{Facultad  de  Ciencias, Universidad  de  la  Rep\'ublica\\
Ig\'ua  4225,  Montevideo,  Uruguay}

\emailAdd{silvia.nagy@durham.ac.uk}
\emailAdd{jperaza@cmat.edu.uy}

\abstract{Working in the self-dual sector for Yang-Mills and gravity, we show how to construct an extended phase space at null infinity, to all orders in the radial expansion. This formalises the symmetry origin of the infrared behaviour of these theories to all subleading orders. As a corollary, we also derive a double copy mapping from a subset of YM gauge transformations to a subset of  diffeomorphisms to all orders in the transformation parameters, which to our knowledge has not been presented before in the literature.}

\begin{document} 
\maketitle
\flushbottom

\section{Introduction}\label{Introduction}

Symmetries for the $S-$matrix have been proven to be a richer family than originally expected.
In the seminal paper \cite{Strominger:2013lka}, the authors showed that asymptotic symmetries provide a Ward identity for the $S-$matrix via the soft photon theorem, later also proved in other contexts \cite{Strominger:2013jfa,He:2014cra}. This equivalence is expected to continue to be valid between sub$^n$-leading theorems \footnote{The term sub$^n$-leading refers to the energy expansion for the factor that relates the amplitude with a soft gauge boson with the same amplitude without the gauge boson.} and new asymptotic symmetries with more general fall offs. These symmetries are not accommodated by the standard symplectic structure of the radiative phase space. In order to put this equivalence on a more solid footing, we are then required to construct an extended phase space with a well-defined variational algebra.
  
For gravity, in \cite{Campiglia2015} it was suggested that the extension of the BMS group should be generalized by taking into account the group $\text{Diff}(S^2)$ for super-rotations, which is the so-called generalised BMS group (GBMS). In \cite{Campiglia2020}, one of the authors constructed charges for GBMS and defined a phase space, such that in this new phase space GBMS acts canonically. The extension is done by considering an extra field supported in the corners of null infinity \footnote{Here \textit{corner} denotes the boundary of the hypersurface on which the symplectic structure is defined.}. Later, in \cite{Campiglia:2021oqz} the same strategy was used to extend the radiative phase space of Yang-Mills theory, obtaining a closed algebra up to linear level in the extension. The charges used come from the leading and subleading charges from the soft theorems (\cite{Strominger2014a},\cite{Lysov:2014csa}) and the phase space was computed so that the charges act canonically. An alternative approach, the corner symmetry construction, was presented originally in \cite{Donnelly:2016auv} and later explored in e.g. \cite{Speranza:2017gxd,Freidel:2020ayo,Freidel:2020svx,Freidel:2020xyx,Freidel:2021dxw,Ciambelli:2021nmv,Freidel:2021cjp,Freidel:2021dfs,Ciambelli:2021vnn}. In contrast with the previous approach, gauge symmetries with any fall off are required to have a vanishing associated charge (pure gauge). The non-trivial charges now come from the corner symmetry.

In the Yang-Mills case, as well as in gravity, it is natural to consider the question of whether it is possible to extend the phase space to account for all the symmetries coming form the sub$^n-$leading/Ward identity equivalences. These symmetries will correspond to symmetry parameters with increasingly divergent behaviour at null infinity coming form the radial expansion. The present work aims to answer this question, at least in a formal approach, by restricting to the self-dual sector (SD) of the theory.

On the other hand, there is a proposed connection between gravity and (two copies of) YM theory, which goes under the name of the double copy\footnote{See \cite{Bern:2019prr,Borsten:2020bgv,Adamo:2022dcm} for some reviews.}. In the self-dual sector this takes a particularly elegant form \cite{Monteiro:2011pc,Chacon:2020fmr,Campiglia:2021srh,Armstrong-Williams:2022apo,Krasnov:2021cva,Elor:2020nqe,Farnsworth:2021wvs,Skvortsov:2022unu,Boels:2013bi,He:2015wgf,Berman:2018hwd,Monteiro:2022nqt}. In the search for a more general double copy map, a valuable criterion is the robustness under local transformations, i.e. the requirement that diffeomorphism transformations in gravity follow from gauge transformations in YM \cite{Borsten:2021zir,Borsten:2020xbt,Luna:2020adi,Borsten:2019prq,Anastasiou:2018rdx,LopesCardoso:2018xes,Anastasiou:2014qba,Borsten:2021hua,Borsten:2020zgj,Bonezzi:2022yuh,Ferrero:2020vww,Godazgar:2022gfw}. In \cite{Campiglia:2021srh} one of the authors constructed the first non-perturbative such map for a subset of the local symmetries, looking at \textit{infinitessimal} transformations. Here, we will be able to extend this to \textit{finite} transformations. We note that this does not follow immediately from the infinitesimal statement, as the exponentiation is different for the gauge field and the graviton. We will also define the map at null infinity\footnote{For other formulations of the double copy at null infinity, including relations to soft theorems, celestial holography and solution-generating transformations see e.g. \cite{BjerrumBohr:2010hn,Oxburgh:2012zr,Naculich:2013xa,He:2014bga,White:2014qia,Vera:2014tda,Luna:2016idw,PV:2019uuv,Monteiro:2022lwm,Adamo:2017nia,Casali:2020vuy,Casali:2020uvr,Kalyanapuram:2020epb,Kalyanapuram:2020aya,Pasterski:2020pdk,Huang:2019cja,Emond:2020lwi,Alawadhi:2019urr,Banerjee:2019saj,Guevara:2022qnm,Godazgar:2021iae,Adamo:2021dfg,Mao:2021kxq}.}, extending the results in \cite{Campiglia:2021srh} to all orders in $r$.

The paper is structured as follows: in \autoref{Light-cone gauge in the self-dual sectors of YM and gravity} we give an overview of the self-dual sectors of gauge theory and gravity in the light-cone formulation. In \autoref{Linear extended phase space} we review the linear phase space extension of \cite{Campiglia:2021oqz}, adapted to the self-dual sector. In \autoref{Full Yang-Mills extension} we give a general procedure for constructing this to all orders in the expansion coordinate for YM, which we adapt in \autoref{Full gravity extension} to gravity as well. As a by-product this gives us a double copy for finite local transformations. In \autoref{extension_Bondi_coord} we take our results to null infinity. In \autoref{An infinite cube of symmetries in the self-dual sector} we comment on the connection to the infinite towers of symmetries present in the self-dual sectors, the various links between them, and their manifestation at null infinity.  We conclude in \autoref{Conclusions}.

\section{Light-cone gauge in the self-dual sectors of YM and gravity}\label{Light-cone gauge in the self-dual sectors of YM and gravity}
We will follow the conventions of \cite{Campiglia:2021srh} and use light-cone coordinates $(U,V,Z,\Zb)$, related to Cartesian coordinates $X^\mu$ via:
\be
U=\frac{X^0-X^3}{\sqrt{2}}, \quad V=\frac{X^0+X^3}{\sqrt{2}}, \quad Z=\frac{X^1+i X^2}{\sqrt{2}}, \quad \Zb =\frac{X^1-i X^2}{\sqrt{2}}. \label{deflccoords}
\ee
It is useful to introduce the notation:
\be
x^i :=(U,\Zb) , \quad y^{\alpha}:=(V,Z) \label{splitcoords}.
\ee
which splits space-time into two $2d$ subspaces. The Minkowski metric is then
\be\label{bkg_metric}
ds^2= 2 \eta_{i \alpha} dx^i dy^\alpha = -2 dU d V+2 d Z d \Zb .
\ee
and we introduce the anti-symmetric "area elements" of the $2d$ subspaces:
\ba \label{omega_and_pi}
 \Omega_{i j} d x^i \wedge d x^j & = & d U  \wedge d \Zb - d \Zb \wedge d U \\ \label{pi_def}
  \Pi_{\alpha \beta}  d y^\alpha \wedge d y^\beta &= & d V \wedge d Z - d Z \wedge d V
\ea
which act as inverses of each other:
\ba
\Omega_i^{\ \alpha} \Pi_{\alpha}^{\ j} = \delta_i^{j} \label{OmegaPi} \\
\Pi_{\alpha}^{\ i} \Omega_{i}^{\ \beta} = \delta_\alpha^{\beta} \label{PiOmega}.
\ea

\subsection{Self-Dual Yang-Mills}
Consider a YM field $\A_\mu$ with field strength 
\be
F_{\mu \nu} = \partial_\mu \A_\nu  - \partial_\nu \A_\mu - i [\A_\mu,\A_\nu],
\ee
The dual is defined as
\be
\tilde{F}_{\mu \nu} := \frac{1}{2}\epsilon_{\mu \nu}^{\phantom{\mu \nu }\rho \sigma} F_{\rho \sigma} . \label{dualF}
\ee
The field strength is self-dual provided 
\be
\tilde{F}_{\mu \nu} = i F_{\mu \nu}. \label{selfdual}
\ee
Then, working in the light-cone gauge $\A_U=0$, the self-dual sector can be described via (see \cite{Campiglia:2021srh,Monteiro:2011pc,Cangemi:1996rx,Chalmers:1996rq,Plebanski:1975wn})
\be
\A_i=0,\quad \mathcal{A}_\alpha =  \Pi_\alpha^{\ i} \partial_i \Phi \label{Aphi}
\ee
with $\Pi_{\alpha\beta}$ as in \eqref{pi_def} and where the Lie algebra valued scalar satisfies
\be 
\square \Phi  =  - i \Pi^{i j}[\partial_i \Phi,\partial_j \Phi]  \label{eomPhi}
\ee
It is easy to see that the Lorenz gauge is satisfied automatically
\be
 \partial^\mu \mathcal{A}_\mu = \eta^{i\alpha}\partial_i\A_\alpha=\Pi^{ij}\partial_i\partial_j\Phi=0
\ee
making use of the anti-symmetry of $\Pi$. It also follows that
\be \label{gauge_Lorenz_cond}
D^\mu \mathcal{A}_\mu =0, \qquad \text{with} \quad D_\mu \A_\nu=\partial_\mu \A_\nu+i[\A_\nu,\A_\mu].
\ee
In this paper, we will be be considering the family of gauge transformations
\be 
\delta_\Lambda \mathcal{A}_\mu =D_\mu\Lambda= \partial_\mu \Lambda + i [\Lambda, \mathcal{A}_\mu], \label{delA}
\ee
which preserve the light-cone gauge condition \eqref{Aphi}\footnote{We are working at linear order in the gauge parameter for now, but we will be extending to all orders in \autoref{Full Yang-Mills extension}.}. It is easy to see that these are parametrised by
\be 
\partial_i \Lambda =0 \implies \Lambda=\Lambda(y) \label{lam}
\ee
We note that this subset of the gauge transformations also automatically preserves the gauge condition \eqref{gauge_Lorenz_cond}. This is one of the crucial simplifications that follow from starting with the more restrictive light-cone gauge, instead of \eqref{gauge_Lorenz_cond}. Had we taken \eqref{gauge_Lorenz_cond} as our starting point and attempted to solve for $\Lambda$ (for example as a perturbation series in the coordinate $V$), we would obtain a field-dependent expression for $\Lambda$ (see \cite{Campiglia:2021oqz} for details), which would have significantly increased the difficulty of calculations in subsequent sections. We leave this extension for future work. 

\subsubsection{Phase space}

In this section, we construct a phase space which will later be extended, first formally in \autoref{Full Yang-Mills extension} and \autoref{Full gravity extension}. Then, in section \ref{extension_Bondi_coord} we will see the implications of such extension in the radiative case, when taking the limit to $\Ib$. 

First, we should restrict to the subset of relevant solutions. This step is in general clear when dealing with boundaries: we take solutions that satisfy certain boundary conditions. In the case of radiative fields, the condition is a sufficiently strong decay of the fields, such that the energy reaching infinity is finite. 

In our case, we take the coordinate $V$ as the expansion parameter, and impose some conditions on the fields which will restrict the phase space. Consider the following expansion for the scalar field\footnote{For simplicity, in this work we do not include logarithmic terms - in our set-up, this is a consistent choice. However, they might appear in more general set-ups, such as in \citep{Campiglia:2021oqz}, where they are required by the consistency of the equations.},

\be \label{Phidecay}
\Phi =\Phi^{(0)}(U,Z,\Zb) + \frac{\Phi^{(-1)}(U,Z,\Zb)}{V} + \frac{\Phi^{(-2)}(U,Z,\Zb)}{V^2} + ...,
\ee 
where the first term is a general term coming from the formula relating $\A$ with $\Phi$, \eqref{Aphi}. 

%
%

We impose the self-duality condition on $\Phi$ \footnote{Since we are working in the SD sector, we are identifying the Bianchi identity with the equation of motion. Therefore, if we have a potential for $\A$, the equations of motion are trivially satisfied.}, given by \eqref{eomPhi}, which, using \eqref{bkg_metric} and \eqref{omega_and_pi} can be written as 

\be \label{kinematicalPhi}
2(\partial_U \partial_V - \partial_Z \partial_{\Zb}) \Phi	= -2 i [\partial_U \Phi , \partial_{\Zb} \Phi],
\ee 
from which we obtain a sequence of equations for each term in the expansion (following from \eqref{Phidecay} and \eqref{kinematicalPhi}),

\beq
\partial_Z \partial_{\Zb} \Phi^{(0)} &= & i[\partial_U \Phi^{(0)}, \partial_{\Zb} \Phi^{(0)}], \\
\partial_Z \partial_{\Zb} \Phi^{(-1)} &=& i[\partial_U \Phi^{(-1)}, \partial_{\Zb} \Phi^{(0)}] + i[\partial_U \Phi^{(0)}, \partial_{\Zb} \Phi^{(-1)}], \label{ord_-1}\\
\partial_Z \partial_{\Zb} \Phi^{(-2)} &=& -\partial_U \Phi^{(-1)} + i[\partial_U \Phi^{(-1)},\partial_{\bar{Z}} \Phi^{(-1)}] + i[\partial_U \Phi^{(0)},\partial_{\bar{Z}} \Phi^{(-2)}]\nonumber\\
&&+ i[\partial_U \Phi^{(-2)},\partial_{\bar{Z}} \Phi^{(0)}] , \label{ord_-2}\\
&...&\label{ord_-n}
\eeq
This set of equations indicates that $\Phi^{(0)}$ can be a candidate for initial data if we take it to be holomorphic,

\be 
\Phi^{(0)} = \Phi^{(0)}(U,Z),
\ee
while the coefficients $\Phi^{(-1)},\Phi^{(-2)},...$ remain to be solved, using \eqref{ord_-1},\eqref{ord_-2},\eqref{ord_-n}.

From the fall-off \eqref{Phidecay}, together with \eqref{Aphi} we have
\ba\label{A_Z_exp}
\A_Z &=& \partial_U \Phi = \A^{(0)}_Z + \frac{\A^{(-1)}_Z}{V} + \frac{\A^{(-2)}_Z}{V^2} + ... \\
\A_V &=& \partial_{\Zb} \Phi = \frac{\A^{(-1)}_V}{V} + \frac{\A^{(-2)}_V}{V^2} + ... \label{A_V_exp}
\ea 
where the coefficients in the expansions are functions of $(U,Z,\Zb)$. Finally, from \eqref{kinematicalPhi} we have several restrictions on the form of the functions in the expansions. We will define our {\it initial phase space} as the set,
\be
\Gamma^{0}_{YM} = \{\Phi^{(0)} \mid \A_i = 0, \A_\alpha = \Pi_\alpha^{\ i} \partial_i \Phi, \text{ with } \Phi^{(0)} \text{ as initial data for $\Phi$.}\}\label{0_th_ph_sp} .
\ee

Next, we study the space of variations and its action on $\Gamma^{0}_{YM}$. Take $\Lambda$ to be the parameter of gauge transformations. Its $V$ expansion is 
\be 
\Lambda = \sum_{n = -\infty}^{+\infty} \Lambda^{(n)}(Z) V^{n},
\ee 
where $\Lambda^{(n)}=\Lambda^{(n)}(Z)$, since the variations must be gauge preserving (see \eqref{lam}). To ensure convergence, for the rest of the paper we will assume that only finitely many of the terms with power $n>0$ are non-zero. Then the variation of $\A_Z$ and $\A_V$ is

\beq
\sum_{n=-\infty}^{\infty} \delta_\Lambda \A_Z^{(n)} V^{n} &=&  \sum_{n = -\infty}^{+\infty} \left( \partial_Z \Lambda^{(n)}- i \sum_{k=0}^\infty \left[\A_Z^{(n-k)} , \Lambda^{(k)} \right] \right) V^{n} \\
\sum_{n=-\infty}^{\infty} \delta_\Lambda \A_V^{(n)} V^{n} &=&  \sum_{n = -\infty}^{+\infty} \left( n  \Lambda^{(n)} V^{-1}- i \sum_{k=0}^\infty \left[\A_V^{(n-k)} , \Lambda^{(k)} \right] \right) V^{n} 
\eeq
Therefore, if we want the gauge transformations to preserve some chosen decay in the fields, we must impose certain conditions on $\Lambda$. The fall-offs for $\A_Z$ and $\A_V$ given in \eqref{A_Z_exp} and \eqref{A_V_exp} are simultaneously preserved by any $\Lambda$ of the form,
\be \label{Lambda_for_0_phase_space}
\Lambda = \sum_{n = - \infty}^{0} \Lambda^{(n)} V^{n},
\ee
where for $\Lambda^{(0)} \neq 0$ corresponds to the large gauge transformations, \cite{Strominger:2013lka}. Unlike in the general case, \cite{Campiglia:2021oqz}, the coefficients $\Lambda^{(n)}$ are field-independent. The action of $\Lambda$ on the scalar is given by \cite{Campiglia:2021srh}
\be 
\delta_\Lambda \Phi  = \Omega^{\ \alpha}_{i} x^i \partial_\alpha \Lambda + i[\Lambda, \Phi],
\ee 
which can be seen to be compatible with the fall off, if $\Lambda$ is of the form  \eqref{Lambda_for_0_phase_space}. 

\subsection{Self-dual Gravity}

Next we turn our attention to gravity. Here, we define the dual of the Riemann tensor as
\be 
\tilde{R}_{\mu \nu \rho}^{\phantom{\mu \nu \rho} \sigma} := \frac{1}{2}\epsilon_{\mu \nu}^{\phantom{\mu \nu }\eta \lambda} R_{\eta \lambda \rho}^{\phantom{\mu \nu \rho} \sigma}
\ee
The self-duality condition is then
\be 
\tilde{R}_{\mu \nu \rho}^{\phantom{\mu \nu \rho} \sigma}  = i R_{\mu \nu \rho}^{\phantom{\mu \nu \rho} \sigma}. \label{dualR}
\ee
It is convenient to split the metric as 
\be 
g_{\mu \nu} = \eta_{\mu \nu}+h_{\mu \nu} \label{defg}
\ee
We emphasize that, despite the notation, we are considering the full non-perturbative metric here. In the light-cone gauge self-dual gravity is then described via
\be \label{hphi_eq}
h_{i\mu}=0,\quad h_{\alpha \beta} = \Pi_\alpha^{\ i} \Pi_\beta^{\ j} \partial_i \partial_j \phi ,
\ee
with the scalar $\phi$ satisfying
\be \label{eomphi}
\square \phi = \frac{1}{2}\Pi^{i j} \Pi^{k l} \partial_i \partial_k \phi \partial_j \partial_l \phi 
= \frac{1}{2}\Pi^{i j} \{\partial_i \phi, \partial_j \phi  \},
\ee
where we introduced the Poisson bracket
\be\label{poisson_def}
 \{f,g\}:= \Pi^{ij}\partial_i f \partial_j g.
\ee
The equation of motion for $\phi$ \eqref{eomphi} is derived in Appendix A of \cite{Campiglia:2021srh} in this notation.

The graviton automatically satisfies a transverse-traceless gauge condition
\be \label{transverse_traceless}
\partial^\mu h_{\mu \nu}=0, \quad \eta^{\mu \nu}h_{\mu \nu }=0 
\ee
We are interested in general coordinate transformations which preserve \eqref{hphi_eq}. At linear order in the parameter, these will be a subset of the diffeomorphism transformations
\be
\delta_\xi h_{\mu \nu} :=  \L_\xi \eta_{\mu \nu} + \L_\xi h_{\mu \nu} \label{delh} 
\ee
parametrised by
\be 
\xi_{i}=0, \quad \xi_\alpha=b_\alpha(y)=\partial_\alpha b(y). \label{xiizero}
\ee
where the final equality ensures that it will also be a symmetry of equation \eqref{eomphi}.We remark that these diffeomorphisms automatically preserve the transverse-traceless gauge condition \eqref{transverse_traceless}.
We note that, unlike in the YM case, this is not the most general symmetry that preserves \eqref{hphi_eq}. These exists a second family of diffeomorphisms, which can be obtained recursively from \eqref{xiizero} (see \cite{Campiglia:2021srh}), and is in fact part of an infinite tower of symmetries arising as a consequence of the integrability of the self-dual sector. We will revisit these in \autoref{An infinite cube of symmetries in the self-dual sector}.

\subsubsection{Phase space}

We take the same fall-off as \eqref{Phidecay} for the gravity scalar, 

\be \label{phidecay}
\phi = \phi^{(0)}(U,Z,\Zb) + \frac{\phi^{(-1)}(U,Z,\Zb) }{V} + ... ,
\ee
which has to satisfy (from \eqref{eomphi}):
\be \label{phi_equation_sd}
-2(\partial_U \partial_V - \partial_Z \partial_{\bar{Z}}) \phi =\frac{1}{2}  \Pi^{ij} \Pi^{kl} \partial_k \partial_i \phi \partial_l \partial_j \phi.
\ee
For $\phi^{(0)}$  this reads 
\be  \label{constraint_phi_0}
\partial_Z \partial_{\bar{Z}} \phi^{(0)} =  \frac{1}{4}\Pi^{ij} \Pi^{kl} \partial_k \partial_i \phi^{(0)} \partial_l \partial_j \phi^{(0)}.
\ee
Equation \eqref{constraint_phi_0} is trivially satisfied if we assume that $\phi^{(0)}$ is holomorphic, and we can take it as a candidate for the initial data, with the other components calculated from \eqref{phi_equation_sd}. Therefore, we define the initial phase space as,
\be\label{0th_ph_sp_gravity}
\Gamma^{0}_{\text{grav}} = \{ \phi^{(0)}  \mid h_{\alpha \beta} =  \Pi_\alpha^{\ i} \Pi_\beta^{\ j} \partial_i \partial_j \phi , \text{ with } \phi^{(0)} \text{ the initial data for } \phi \}.
\ee
The diffeomorphism with parameter \eqref{xiizero} will act on the scalar $\phi$ as
\be 
\delta_b \phi = \Omega_i^{\ \alpha} \Omega_{j}^{\ \beta} x^i x^j \partial_\alpha \partial_\beta b + \eta^{i \alpha} \partial_\alpha b \partial_i \phi,
\ee
By defining the ``Hamiltonian'' of the vector field $\xi$ as \cite{Campiglia:2021srh},
\be \label{ham_def}
\lambda := 2 \Omega^{\ \alpha}_i x^i \xi_\alpha= 2 \Omega^{\ \alpha}_i x^i \partial_\alpha b (y), 
\ee
we can write the variation of $\phi$ as
\be \label{delta_phi_grav}
\delta_b \phi = \frac{1}{2} \Omega^{\ \alpha}_i x^i \partial_\alpha \lambda - \frac{1}{2} \{\lambda , \phi\}.
\ee
We note the useful identity
\be \label{useful_poisson_xi}
\xi^i\partial_i=-\frac{1}{2}\{\lambda(\xi),\cdot\} 
\ee
when $\xi$ is defined as in \eqref{xiizero}.
The fall-off of the diffeomorphism has to be consistent with the definition of $\Gamma^{(0)}_{\text{grav}}$, as in the SDYM case above. It can be shown that the diffeomorphism generator is given by
\be 
\lambda = \sum_{n=-\infty}^{0} V^n \lambda^{(n)}.
\ee

\subsubsection{Double copy}\label{double_copy_old}

The similarities between the scalar equations \eqref{eomPhi} and \eqref{eomphi}  suggest simple double-copy replacement rules \cite{Monteiro:2011pc}:
\be \label{color_kin_eom}
 \Phi \to \phi, \qquad -i[ \ , \ ] \to \tfrac{1}{2} \{ \ , \ \}
\ee
and the fields $\A_\alpha$ and $h_{\alpha\beta}$ are also directly mapped via the operator $\Pi_\beta^{\ j} \partial_j$. The residual symmetry transformations also obey a double copy relation. The graviton transformation can be written in terms of the Hamiltonian \eqref{ham_def} as
\be
\delta_\lambda h_{\alpha\beta}  =  \Pi_{(\alpha}^{\ i}\partial_i \left( \partial_{\beta)} \lambda - \tfrac{1}{2}\Pi_{\beta)}^{\ j}\partial_j\{\lambda,\phi \}\right),
\ee
This is immediately obtained from the YM transformation, written below in terms of the scalar $\Phi$
\be 
\delta_\Lambda \A_\alpha  =  \partial_\alpha\Lambda +  i \Pi_\alpha^{\ i}\partial_i [\Lambda,  \Phi]  \\
\ee
by making use of the double copy rules \eqref{color_kin_eom} , supplemented by
\be \label{gauge_to_hamiltonian}
\Lambda \to \lambda
\ee
In the next few sections, we will see that these double copy relations continue to hold for the extended phase spaces that we are constructing.

\section{Linear extended phase space} \label{Linear extended phase space}
\subsection{Yang-Mills}\label{YM linear phase space}
Here we review the linear extended phase space construction of \cite{Campiglia:2021oqz}, specialising to the subset of symmetries in the self-dual sector which we described in \autoref{Light-cone gauge in the self-dual sectors of YM and gravity}. In \cite{Campiglia:2021oqz}, one of the authors implemented the null infinity limit in Bondi retarded coordinates, $(\bar{r},\bar{u},z,\bar{z})$, and took the order $\bar{r}^0$ and $\bar{r}^1$ large gauge transformations (LGT) in a linearly extended phase space $\Gamma^1_{YM}$.

In our case, we will initially work in light-cone coordinates and consider the fall-off in the coordinate $V$. As we will see in \autoref{extension_Bondi_coord}, this expansion is analogous to the fall-off in $r$ when going to null infinity, see \eqref{Bondi_coordinates}. As explained in \autoref{Introduction}, we want to allow for a gauge generator with arbitrary higher powers of $V$ in the fall-off,
\be\label{Lambda_expansion}
\Lambda = \sum_{n=-\infty}^{+\infty} V^n \Lambda^{(n)},
\ee
where $\Lambda^{(n)}=\Lambda^{(n)}(Z)$. This of course violates the fall-offs in \eqref{A_Z_exp} and \eqref{A_V_exp}. The first step in resolving this issue was taken in \cite{Campiglia:2021oqz}, where a linear extension of the field was given. To show how one arrives at this result, we first attempt to define the extended phase space candidate to all orders
\be \label{ext_space_lin_YM}
\Gamma^{\text{ext}}_{\text{YM}}:= \Gamma^{0}_{YM} \times \left\lbrace \Psi \mid  \quad \Psi = \sum_{n=1}^{+\infty} V^n \Psi^{(n)} (Z) \right\rbrace,
\ee 
with $\Gamma^{0}_{YM}$ as in \eqref{0_th_ph_sp}, and each pair $(\A_\alpha(\Phi^{(0)}),\Psi)$ in the Cartesian product above defines a \textit{new} vector potential in the linear extension, 
\be \label{A_ext_lin}
\tilde{\A}_\alpha := \A_\alpha +D_\alpha \Psi
\ee
with $\A_\alpha$ as in \eqref{Aphi}. In this way, we are splitting the original full phase space in the bulk (where $\A_\alpha$ has any arbitrary $V^n$ order), into two contributions, the $\Gamma^{0}_{YM}$ part plus a LGT generated part.\footnote{These types of subsystems and extensions have been introduced in \cite{Donnelly:2016auv} for the case of Yang-Mills and gravity in the presence of boundaries.} We will show below that \eqref{A_ext_lin} forces us to set $\Psi^{(n)}(Z)= 0$ for $n \geq 2$.

We consider any LGT, as in \eqref{Lambda_expansion}, which can be decomposed as,
\be \label{splitting_Lambda}
\Lambda = \sum_{n=-\infty}^{0} V^n \Lambda^{(n)} + \sum_{n=1}^{+\infty} V^n \Lambda^{(n)} =: \Lambda_{-} + \Lambda_{+}.
\ee
We will derive an expression for $\delta_{\Lambda} \Psi$ using the consistency of the gauge condition, i.e., \emph{the extended gauge field $\tilde{\A}_\alpha$ transforms as a gauge field in the extended space},
\be  \label{deformgauge}
\delta_\Lambda \tilde{\Av}_\alpha = \tilde{D}_\alpha \Lambda,
\ee
where $\tilde{D}$ is the covariant derivative associated to $\tilde{\Av}$. This is the main identity that must be preserved in any extension supported on the bulk \cite{Campiglia:2021oqz}.

By direct computation, \eqref{deformgauge} gives
\be\label{cons_lin_YM}
\begin{aligned}
\delta_\Lambda \A_\alpha -i [\delta_\Lambda \A_\alpha , \Psi] + D_\alpha \delta_\Lambda \Psi &= \partial_\alpha \Lambda -i [\A_\alpha , \Lambda] -i [\partial_\alpha \Psi, \Lambda] - [[\Av_\alpha ,\Psi], \Lambda] \\
&= -i[D_\alpha \Lambda ,\Psi] + D_\alpha \left(\Lambda -i [\Psi, \Lambda] \right).
\end{aligned}
\ee

In order to solve it, we have to prescribe the action of $\Lambda_{\pm}$ on $\A$, and deduce its action on $\Psi$. We present the prescription in analogy with what was done in \cite{Campiglia:2021oqz}.

At this stage we can focus on $A^{(0)}_Z$, which will correspond to the non-vanishing component when going to null infinity. The covariant derivative at $\mathcal{O}(V^0)$ acts as,
\be 
D^{(0)}_Z X = \partial_Z X - i [A^{(0)}_Z , X].
\ee
We \textit{define} the action of $\Lambda_{\pm}$ on $\A_Z$ as,

\be \label{variations_Lambda_on_A}
\delta_{\Lambda_{-}} \A_Z = D_Z \Lambda_{-}, \quad \delta_{\Lambda_{+}} \A_Z = 0,
\ee
where the variation of $\A_Z^{(0)}$ w.r.t. $\Lambda$ can be read off as 
\be\label{variation_A_Z_0}
\delta_\Lambda A^{(0)}_Z = \partial_Z \Lambda^{(0)} - i [\A_Z^{(0)} , \Lambda^{(0)}].
\ee
Here the expected extra terms $i\sum_{k\geq 1}[ A^{(-k)}_Z , \Lambda^{(k)}]$ do not enter the definition, due to the projection \eqref{variations_Lambda_on_A}. As we will see below, we can incorporate them in the variation $\delta_\Lambda \Psi$. 

From \eqref{cons_lin_YM}, focusing on $\alpha=Z$, splitting $\Lambda$ as in \eqref{splitting_Lambda}, and using the definition \eqref{variations_Lambda_on_A}, we arrive at
\be\label{cons_lin_YM_simplified}
D_Z \delta_{\Lambda} \Psi = -i[D_Z \Lambda_{+} ,\Psi] + D_Z \left(\Lambda_{+ } -i [\Psi, \Lambda] \right).
\ee
We show in \autoref{Some_YM_calcs} that the above can be solved when working to linear order in $V$ (and in the fluctuations $\Psi^{(n)},\Lambda_+$), to give:
\be \label{delta_psi_1_lins}
(\delta_{\Lambda} \Psi)^{(1)} =\Lambda_{+}^{(1)}-i[\Psi^{(1)}, \Lambda_-^{(0)}]
\ee
We will see in \autoref{Full Yang-Mills extension} that in order to accommodate a fall-off of $\mathcal{O}(V^n)$ (corresponding to $\mathcal{O}(r^n)$ at null infinity), we will need to further extend the phase space \eqref{ext_space_lin_YM}-\eqref{A_ext_lin} to $\tilde{\A}_\alpha=\tilde{\A}_\alpha(\A_\alpha,\Psi,...,\Psi^n)$.

For now, we have to reduce the phase space extension, and only consider $\Psi$ fields of $\mathcal{O}(V^1)$, i.e., 

\be \label{lin_ph_sp}
\Gamma^{\text{ext}}_{\text{lin,YM}} = \Gamma^{0}_{YM} \times \left\lbrace \Psi \mid  \quad \Psi = V \Psi^{(1)} (Z) \right\rbrace,
\ee
and therefore the extension can only be linear. Regarding the algebra of variations, since the linearized extension is valid only up to order $O(V^1)$, we have to check the commutators of $[\delta_{\Lambda_{1,\pm}} , \delta_{\Lambda_{2,\pm}} ]$ on the phase space, for $\Lambda_{+ }$ of order $V$ and $\Lambda_{-}$ acting on $A^{(0)}_Z$. To simplify notation, we will call $\Lambda_{+}$ as $\Lambda^1$ and $\Lambda_{-}$ as $\Lambda^0$, in analogy with \cite{Campiglia:2021oqz},

\begin{eqnarray} \label{linearised_commutators_1}
\left[\delta_{\Lambda^0_1} , \delta_{\Lambda^0_2}\right] &=&- i\delta_{[\Lambda^0_1 , \Lambda^0_2]},\\ \label{linearised_commutators_2}
\left[\delta_{\Lambda^1_1} , \delta_{\Lambda^0_2}\right] &=&- i\delta_{[\Lambda^1_1 , \Lambda^0_2]}, \\ \label{linearised_commutators_3}
\left[\delta_{\Lambda^1_1} , \delta_{\Lambda^1_2}\right] &=& 0, 
\end{eqnarray}
where the last bracket vanishes in the linearized (in $\Lambda^1$) approximation. The commutators above are inherited from the commutators corresponding to the transformation \eqref{deformgauge}, using \eqref{A_ext_lin}, indeed 
\be
\left[\delta_{\Lambda_1}, \delta_{\Lambda_2}\right]\tilde{\Av}_\alpha = i\delta_{[\Lambda_1,\Lambda_2]}\tilde{\Av}_\alpha
\ee

This construction is related to the introduction of the ``extra'' field in \cite{Donnelly:2016auv}, which is a local trivialization  of the bundle around the corner $S = \partial \Sigma$, where $\Sigma$ is the hypersurface on which the symplectic structure is computed. This new field takes into account the dynamics of the boundary. In \cite{Donnelly:2016auv,Ciambelli:2021nmv,Speranza:2017gxd} there are similar formulas for the embedding map in the case of gravity.

It is of course not a coincidence that $\tilde{\A}_\alpha$ has the functional form of a gauge transformation of $\A_\alpha$, with the gauge parameter replaced by the field $\Psi$. This is reminiscent of the St\"uckelberg procedure\footnote{Initially introduced in the context of electromagnetism \cite{Stueckelberg:1938hvi}, also extended to (super)gravity \cite{Kuchar:1991xd,Nagy:2019ywi}.}, which reinstates a broken local symmetry in the action of some field theory via the introduction of additional fields transforming non-linearly (similar to our $\Psi$). The St\"uckelberg procedure consists of performing a transformation of the non-invariant action (for example a gauge theory with a mass term), and then promoting the parameters to new fields. 

If we wish to extend our phase space to $\mathcal{O}(V^n)$ (corresponding to $\mathcal{O}(\Psi^n)$), the natural guess is then that $\tilde{\A}_\alpha$ should functionally look like the full gauge transformation of $\A_\alpha$ (i.e. the exponentiated group action), truncated to order $n$. As we will see in \autoref{Full Yang-Mills extension}, this indeed turns out to be the case.  

\subsection{Gravity}

We will proceed along the same lines as in the YM case. Completely analogous arguments to the YM case prohibit the construction of the phase space beyond $\mathcal{O}(V^1)$ in this approach, so in this section we will just look for a transformation parameter with fall-off
\be \label{lambda_epannsion}
\lambda = \sum_{n=-\infty}^{1} V^n \lambda^{(n)}  
\ee
where $\lambda$ is the "Hamiltonian" \eqref{ham_def} built out of the diffeomophism parameter $\xi^\mu$ given in \eqref{xiizero}. This will violate the fall-off in \eqref{phidecay}. To rectify this, we construct the linear extension of the phase space as
\be\label{ext_phase_lin_grav}
\Gamma^{\text{ext}}_{\text{grav}}:= \Gamma^{0}_{\text{grav}} \times \left\lbrace \psi \mid  \quad \psi =  V \psi^{(1)} (Z) \right\rbrace,
\ee
with $\Gamma^{0}_{\text{grav}}$ as in \eqref{0th_ph_sp_gravity} and each pair $(h_{\alpha\beta},\psi)$ in the Cartesian product above defining a \textit{new} graviton in the linear extension: 
\be 
\tilde{h}_{\alpha\beta}=h_{\alpha\beta}+ \Pi_{(\alpha}^i \partial_i \partial_{\beta)} \psi - \frac{1}{2} \{ \psi ,h_{\alpha\beta} \}
\ee
We note that $\tilde{h}_{\alpha\beta}$ is related to $h_{\alpha\beta}$ via a diffeomorphism transformation (see \autoref{Restricted diffeomorphism transformation}), where the "Hamiltonian" transformation parameter $\lambda$ has been replaced by $\psi$. This is of course in complete analogy with the YM construction in \autoref{YM linear phase space}. The form of $\psi$ is then inherited from that of $\lambda$ \eqref{ham_def}, and we have
\be
 \psi^{(1)}(x^i,Z)=2 \Omega_i^{\ \alpha} x^i \partial_\alpha\beta^{(1)},
\ee
with $\beta^{(1)}=\beta^{(1)}(Z)$. The consistency condition for the gravity extended phase space is\footnote{Following the same logic as in \autoref{Restricted diffeomorphism transformation}, with $h_{\alpha\beta}\to\tilde{h}_{\alpha\beta}$.}
\be \label{grav_linear_consistency}
\delta_\lambda\tilde{h}_{\alpha\beta}=\mathcal{L}_{\lambda(\xi)}\tilde{g}_{\alpha\beta}= \Pi_{(\alpha}^i \partial_i \partial_{\beta )} \lambda - \frac{1}{2} \{ \lambda ,\tilde{h}_{\alpha \beta} \},
\ee
i.e. we want $\tilde{h}$ to transform as a graviton in the extended space. We will now focus on the $\alpha\beta=ZZ$ component of the equation above (this will give the asymptotic shear $C_{zz}$ which captures the free data at null infinity). 
Then, working perturbatively in $V$ and, as in the YM case, requiring that $\lambda^{(1)}$ does not transform $h_{ZZ}$, we have
\be
\delta_\lambda h_{ZZ}^{0}=\Pi_{Z}^i \partial_i \partial_{Z} \lambda^{(0)} - \frac{1}{2} \{ \lambda^{(0)} ,h^{(0)}_{ZZ} \}
\ee
and 
\be \label{delta_psi_1}
\delta_\lambda \psi^{(1)} = \lambda^{(1)} +\tfrac{1}{2} \{\psi^{(1)},\lambda^{(0)}\}.  
\ee
Finally, the linearised commutators \eqref{linearised_commutators_1}-\eqref{linearised_commutators_3} will now have completely analogous forms in gravity in terms of the Poisson brackets.

As mentioned above, higher orders in $V$ will not work in this approach, so one has to extend the phase space to higher orders in $\psi$. We will do this by exponentiating the action of the Lie derivative in \autoref{Full gravity extension}.\\ \\
\noindent

\textbf{Double copy relations:} 
The double copy for the self-dual fields and their symmetry transformations has already been established \cite{Monteiro:2011pc,Campiglia:2021srh}, as described towards the end of \autoref{Light-cone gauge in the self-dual sectors of YM and gravity}. It remains to extend it to the new scalars $\Psi$ and $\psi$, and their symmetry transformations \eqref{delta_psi_1_lins} and \eqref{delta_psi_1}. This is very straightforward, as we notice that the double copy replacement rules \eqref{color_kin_eom} and \eqref{gauge_to_hamiltonian} simply need to be supplemented with
\be \label{dc_psi_1}
\Psi^{(1)} \to \psi^{(1)}
\ee 
For now, we are only able to map the coefficients of $V^1$ in the respective expansions of $\Psi$ and $\psi$. We will extend this to all orders in \autoref{Double copy for full extensions}.

\section{Yang-Mills extension to all orders}\label{Full Yang-Mills extension}
We propose an extension of \eqref{lin_ph_sp} and \eqref{A_ext_lin} to all orders in $\Psi$, given by

\be \label{ext_space_full_YM}
\Gamma^{\text{ext}}_{\infty,\text{YM}}:=\Gamma^0_{YM} \times \{\Psi \mid \Psi = \sum_{n=1}^{+\infty} V^n \Psi^{(n)} (Z) \}.
\ee 
such that
\be \label{A_tilde_full}
\hat{\A}_\alpha = e^{i\Psi} \Av_\alpha e^{-i\Psi} + ie^{i\Psi} \partial_\alpha e^{-i\Psi},
\ee
where $\A_\alpha \in \Gamma^{0}_{\text{YM}}$, and the coefficients $\{\Psi^{(n)} (Z) \}_{n \geq 1}$ are taken such that only finitely many of them are non-zero. This is the natural generalisation of \eqref{A_ext_lin}, as it takes the functional form of a full gauge transformation, with the parameter replaced by the field $\Psi$.

The consistency condition \eqref{deformgauge} now becomes
\be \label{consistency_full_YM}
\delta_\Lambda \hat{\Av}_\alpha = \hat{D}_\alpha \Lambda.
\ee
Note that it is still sufficient to work to linear order in the transformation parameter $\Lambda$\footnote{In principle, we could consider these transformations to all orders in $\Lambda$, however this is not necessary for applications to soft theorems.}. Then, making use of 
\be\label{der_exp_map}
\delta e^X=e^X \Op_X (\delta X)
\ee
with
\be \label{O_operator_def}
\Op_X :=  \frac{1 - e^{-ad_X}}{ad_X} = \sum_{k=0}^{\infty}\tfrac{(-1)^k}{(k+1)!} \left( ad_{X} \right)^k, 
\ee
where $ad_X(Y)=[X,Y]$, we can rewrite \eqref{consistency_full_YM} as (see \autoref{sandwich} for details)

%

\be \label{rearranged_YM_full_cond}
e^{i\Psi}\left[\delta A_\alpha + D_\alpha  \left( e^{-i\Psi} \Op_{-i\Psi}(\delta_\Lambda \Psi)e^{i\Psi} \right)  \right]e^{-i\Psi}
 =e^{i\Psi}\left[ D_\alpha (e^{-i\Psi} \Lambda e^{i\Psi})  \right]e^{-i\Psi}.
\ee
 
This equation, as it stands, is valid everywhere. As in \autoref{YM linear phase space}, we need to write \eqref{rearranged_YM_full_cond} in the phase space variables, and prescribe the action of the variation on $A_Z^{(0)}$. We will define the same transformation rule for $A_Z^{(0)}$ as in equation \eqref{variation_A_Z_0}:
\be
\label{A_Z_var_in_full}
\delta_\Lambda A^{(0)}_Z = D_Z^{(0)} \Lambda^{(0)} 
\ee
where the expansion of $\Lambda$ is given in \eqref{Lambda_expansion}. The interesting part follows when considering the variation of $\Psi$, which turns out to satisfy a remarkably simple constraint, at all orders in $V$, as can be read off directly from \eqref{rearranged_YM_full_cond}, upon substracting \eqref{A_Z_var_in_full}:
\be
\Op_{-i\Psi}(\delta_\Lambda \Psi) = \Lambda -e^{i\Psi}\Lambda^{(0)}e^{-i\Psi} . 
\ee
Formally, we can write
\be \label{delta_Psi_formal}
\left[ \delta_\Lambda \Psi\right]^{(n)} = \left( \Op^{-1}_{-i\Psi}(\Lambda - e^{i\Psi}\Lambda^{(0)}e^{-i\Psi}) \right)^{(n)}.
\ee
where $^{(n)}$ denotes the coefficient of $V^n$ in the $V-$expansion. We show below that the operator $\Op$ is indeed invertible, working order by order in $\Psi$. Let $\delta^{[m]}X$ denote the variation of $X$ at order $m$ in $\Psi$, and $T^{[m]}$ the component of order $\Psi^m$ in the expression $T$.\footnote{This is to distinguish it from the expansion in $V$, which we have denoted by $T^{(n)}$.} Then, in equation \eqref{delta_Psi_formal} we have two expansions, one in $V$ and on in $\Psi$, where each equation reads, 

\be \label{delta_Psi_formal_double}
\left[ \delta^{[m]}_\Lambda \Psi\right]^{(n)} = \left( \Op^{-1}_{-i\Psi}(\Lambda - e^{i\Psi}\Lambda^{(0)}e^{-i\Psi}) \right)^{[m],(n)}.
\ee

First, observe that since $\Psi$ is of orders $V^1$ and higher, then the equations in \eqref{delta_Psi_formal_double} are meaningful only when $n \geq m$, since a $\Psi^m$ term starts at order $V^m$. Then, we have a natural ``induction step'' to solve the inverse operator both in $V$ and in $\Psi$. By the definition of $\Op_{- i \Psi}$ we have
\be
\begin{aligned}
\Op_{-i\Psi}(\delta_\Lambda \Psi)=&\sum_{k=0}^{\infty}\tfrac{(-1)^k}{(k+1)!} \left( ad_{-i\Psi} \right)^k (\delta_\Lambda \Psi)\\
=& \delta_\Lambda \Psi + \frac{i}{2} [\Psi ,\delta_\Lambda \Psi] - \frac{1}{6} [\Psi ,[\Psi ,\delta_\Lambda \Psi]] - \frac{i}{24} [\Psi ,[\Psi ,[\Psi ,\delta_\Lambda \Psi]]] +...,
\end{aligned}
\ee
 Then for $[n]=0$ 
\be \label{0_th_order_YM_from_full}
\delta^{[0]}_\Lambda \Psi  = \Lambda -\Lambda^{(0)}
\ee
This is true to all orders in $V$. When restricting to linear order in V, we find agreement with the first term in \eqref{delta_psi_1_lins}, as expected, using the expansions for $\Lambda$ and $\Psi$ given in \eqref{Lambda_expansion} and \eqref{ext_space_full_YM}. For $[n]=1$, we get
\be
\delta^{[1]}_\Lambda \Psi + \frac{i}{2} [\Psi, \delta^{[0]}_\Lambda \Psi]=-i[\Psi,\Lambda^{(0)}] \\
\ee
Making use of \eqref{0_th_order_YM_from_full}, this becomes
\be \label{1_st_order_YM_from_full}
\delta^{[1]}_\Lambda \Psi =-\tfrac{i}{2}[\Psi,\Lambda+\Lambda^{(0)}]
\ee
which, to linear order in $V$, agrees with the second term in \eqref{delta_psi_1_lins} as expected, again using \eqref{Lambda_expansion} and \eqref{lin_ph_sp}. We introduce the notation $\overset{n}{\delta}$ to denote the variation \textit{up to} order $n$ in $\Psi$. Then
\be 
\overset{1}{\delta}\Psi := \delta^{[0]}_\Lambda \Psi  +\delta^{[1]}_\Lambda \Psi  =\Lambda -\Lambda^{(0)}-\tfrac{i}{2}[\Psi,\Lambda+\Lambda^{(0)}]
\ee

 Now we can press on to higher orders. For example, at $[n]=2$ we have 
\be
\delta^{[2]}_\Lambda \Psi +  \frac{i}{2} [\Psi, \delta^{[1]}_\Lambda \Psi] - \frac{1}{6} [\Psi , [\Psi , \delta^{[0]}_\Lambda \Psi]]
=\tfrac{1}{2}[\Psi,[\Psi,\Lambda^{(0)}]]
\ee
Then, plugging in \eqref{0_th_order_YM_from_full} and \eqref{1_st_order_YM_from_full}, we get
\be
\delta^{[2]}_\Lambda \Psi= -\frac{1}{12} [\Psi , [\Psi , \Lambda-\Lambda^{(0)} ]] 
\ee
This will allow us to extract the transformation of $\Psi$ at $\mathcal{O}(V^2)$. 


It turns out that we can explicitly solve \eqref{delta_Psi_formal_double} in order to give an expression for the variation of $\Psi$ at order n. We use the expansion
\be
\frac{x}{1 - e^{-x}}  = \sum_{m=0}^\infty \frac{B^+_m x^m}{m!},
\ee
where $B^+_n$ are the Bernoulli numbers
\be
B^+_0=1,\ B^+_1=\frac{1}{2},\ B^+_2=\frac{1}{6},\ ...
\ee
One can then show that (see \autoref{Extended phase space transformation at all orders} for details):
\be \label{solved_transform_all_orders}
\delta^{[m]}_\Lambda \Psi=\frac{B_m^+}{m!}\left(ad_{-i\Psi}\right)^m\left[\Lambda+(-1+2\delta_{m,1})\Lambda^{(0)} \right]
\ee
We can now use this to extract the transformation at any order in $V$.

\section{Full gravity extension}\label{Full gravity extension}
In order to proceed to the full gravity extension, we first recall that the diffeomorphism transformation to all orders in the parameter can be written as 
\be 
g'_{\mu\nu} =\left( e^{\Li_\xi} g\right)_{\mu\nu}
\ee
with the Lie derivative 
\be
(\Li_\xi g)_{\mu\nu}=\xi^\rho\partial_\rho g_{\mu\nu}+2\partial_{(\mu}\xi^\rho g_{\nu)\rho} 
\ee
However, recall that we are only interested in a subset of diffeomorphisms parametrised by \eqref{xiizero}. Additionally, we have found it useful to introduce the "Hamiltonian" quantity $\lambda=\lambda(\xi)$ \eqref{ham_def}, which allowed us to write (see \autoref{Restricted diffeomorphism transformation}):
\be
(\Li_\xi g)_{\alpha\beta} =\Pi_{(\alpha}^i \partial_i \partial_{\beta )} \lambda - \frac{1}{2} \{ \lambda ,h_{\alpha \beta} \}
\ee
Let us denote the Poisson bracket \eqref{poisson_def} by
\be \label{pb_notation}
\pb_\theta := \{\theta , \cdot \}
\ee
Then one can show that (see \autoref{Full gravity calculations})
\be
(\Li_\xi) ^{n} g_{\alpha \beta} = \frac{1}{(-2)^n} \pb^n_\lambda (h_{\alpha \beta}),\qquad n>1
\ee
 which allows us to write the all orders transformation rule as
\be
g'_{\alpha\beta} =\Pi_{(\alpha}^i \partial_i \partial_{\beta )} \lambda + e^{-\frac{1}{2} \pb_\lambda} g_{\alpha \beta}
\ee
We now define our extended phase space as 
\be \label{ext_space_full_gravity}
\Gamma^{\text{ext}}_{\infty,\text{grav}}:=\Gamma^0_{grav} \times \{\psi \mid \psi = \sum_{n=1}^{+\infty} V^n \psi^{(n)} (Z) \}.
\ee 
such that the extended graviton is defined as
\be 
\hat{g}_{\alpha\beta}=\Pi_{(\alpha}^i \partial_i \partial_{\beta )} \psi + e^{-\frac{1}{2} \pb_\psi} g_{\alpha \beta}
\ee
where $g_{\alpha\beta}\in \Gamma^0_{grav}$ and the extended phase space fields are of the form
\be \label{psi_n_lim_xi}
 \psi^{(n)}(x^i,Z)=2 \Omega_i^{\ \alpha} x^i \partial_\alpha\beta^{(n)}(Z).
\ee
Note that the definition above still preserves the splitting \eqref{defg} and we have
\be 
\hat{g}_{\mu\nu}=\eta_{\mu\nu}+\hat{h}_{\mu\nu}
\ee 
with $\hat{h}_{i\mu}=0$ and
\be \label{hat_h_with_exp}
\hat{h}_{\alpha\beta}=\Pi_{(\alpha}^i \partial_i \partial_{\beta )} \psi + e^{-\frac{1}{2} \pb_\psi} h_{\alpha \beta}
\ee
The consistency condition \eqref{grav_linear_consistency} is generalised to
\be\label{consistency_cond_grav_full}
\delta \hat{h}_{\alpha\beta}=\mathcal{L}_{\lambda(\xi)}\hat{g}_{\alpha\beta}= \Pi_{(\alpha}^i \partial_i \partial_{\beta )} \lambda - \frac{1}{2}  \pb_\lambda \hat{h}_{\alpha \beta} 
\ee
where, as in the YM case, we found it sufficient to work to linear order in the transformation parameter. At this stage, we could proceed as in the YM case and solve the consistency condition above in order to read off the transformation rule for $\psi$. However, we find it both simpler and more instructive to obtain this from the double copy rules in the next section.

\subsection{Double copy for all order extensions}\label{Double copy for full extensions}
We will show that, just like in the case of the linear extensions, the fully extended phase spaces for YM and gravity are related by the double copy rules described at the end of \autoref{Light-cone gauge in the self-dual sectors of YM and gravity}. To see this, we first note that, using \eqref{exp_identity_comm} and \eqref{der_exp_map}
we can rewrite the extended vector potential \eqref{A_tilde_full} as
\be \label{YMext}
\hat{\Av}_\alpha = e^{i ad_\Psi} A_\alpha (\Phi) + \Op_{-i\Psi} ( \partial_\alpha \Psi) ,
\ee
We will now make use of the double copy rules in \autoref{double_copy_old}. With our new notation, the second term in \eqref{color_kin_eom} becomes
\be\label{ad_to_ad}
-i ad  \to \tfrac{1}{2} \pb
\ee
We will also need \eqref{gauge_to_hamiltonian}, as well as the natural extension of \eqref{dc_psi_1} to all orders in $V$, 
\be\label{DC_Psi_order_n}
\Psi^{(n)} \to \psi^{(n)} ,\quad n\geq1 \ .
\ee
We propose that the double copy of $\hat{A}_\alpha$ is \footnote{We have used the notation $A_\alpha(\phi)=\Pi_\alpha^{\ i} \partial_i \phi$. In order to commute $\Pi_{(\alpha}^{\ i}\partial_i $ and $e^{-\tfrac{1}{2}\pb_\psi}$, we used the fact that $\psi$ is linear in $x^i$, and hence $e^{-\tfrac{1}{2}\pb_\psi}$ is independent of $x^i$.}
\be
\begin{aligned}\label{H_hat_from_DC}
\hat{H}_{\alpha\beta}=& \Pi_{(\alpha}^{\ i}\partial_i \left( e^{-\tfrac{1}{2}\pb_\psi} A_{\beta)}(\phi) + \mathcal{W}_{\tfrac{1}{2}\psi}  ( \partial_{\beta)} \psi)\right)\\
=& e^{-\tfrac{1}{2}\pb_\psi} h_{\alpha\beta} + \Pi_{(\alpha}^{\ i}\partial_i \left(\mathcal{W}_{\tfrac{1}{2}\psi} ( \partial_{\beta)} \psi)\right)
\end{aligned}
\ee
where $\mathcal{W}_{\tfrac{1}{2}\psi} $ is the double copy of $\mathcal{O}_{-i\Psi}$ under the rules \eqref{color_kin_eom}, i.e. 
\be \label{DC_O_W}
\Op_{-i\Psi} :=  \frac{1 - e^{-ad_{-i\Psi}}}{ad_{-i\Psi}}\quad\to\quad 
\mathcal{W}_{\tfrac{1}{2}\psi}:=  \frac{1 - e^{-\pb_{\tfrac{1}{2}\psi}}}{\pb_{\tfrac{1}{2}\psi}}
=\sum_{k=0}^{\infty}\tfrac{(-1)^k}{(k+1)!} \left( \pb_{\tfrac{1}{2}\psi}\right)^k
\ee
Plugging this back into \eqref{H_hat_from_DC} and using the commutativity of the partial derivatives, we get
\be 
\begin{aligned}
\hat{H}_{\alpha\beta}=& e^{-\tfrac{1}{2}\pb_\psi} h_{\alpha\beta} + \mathcal{W}_{\tfrac{1}{2}\psi} (\Pi_{(\alpha}^{\ i}\partial_i  \partial_{\beta)} \psi)\\
=& e^{-\tfrac{1}{2}\pb_\psi} h_{\alpha\beta} +\Pi_{(\alpha}^{\ i}\partial_i  \partial_{\beta)} \psi\\
=&\hat{h}_{\alpha\beta}
\end{aligned}
\ee
thus establishing the double copy - see \eqref{hat_h_with_exp}! In the above, to get to the second line, we used the fact that only the $k=0$ term in \eqref{DC_O_W} contributes, due to the fact that $\psi$ is linear in $x^i$ (\eqref{psi_n_lim_xi})  and remembering the definition of the Poisson bracket \eqref{pb_notation} and \eqref{poisson_def}.

Recall that $\hat{h}_{\alpha\beta}$ and $\hat{A}_\alpha$ formally look like diffeomorphism transformations and, respectively, gauge transformations to all orders, with parameters replaced by $\psi$ and $\Psi$. We thus have a double copy for (a subset of) local symmetries to all orders in the parameters, which, to our knowledge, has not been previously presented in the literature (the results in \cite{Campiglia:2021srh}, while non-perturbative in the fields, were still linear in the transformation parameters). To summarise our results in the more standard language, we have found that 
\be 
\A'_\alpha = e^{i\Lambda} \Av_\alpha e^{-i\Lambda} + ie^{i\Lambda} \partial_\alpha e^{-i\Lambda},
\ee
with
\be
\Lambda=\Lambda(y)
\ee
double copies to
\be
g'_{\mu\nu} =\left( e^{\Li_\xi} g\right)_{\mu\nu}
\ee
with
\be 
\xi_{i}=0, \quad \xi_\alpha=b_\alpha(y)=\partial_\alpha b(y). 
\ee
Finally we can also use the double copy to read off the transformation of $\psi$ to arbitrary orders in $V$ and $\psi$. Starting with \eqref{solved_transform_all_orders} and applying the double copy rules (specifically, we need \eqref{gauge_to_hamiltonian}, \eqref{ad_to_ad}, and \eqref{DC_Psi_order_n}), we get
\be
\delta^{[m]}_\lambda \psi=\frac{B_m^+}{m!}\left(\tfrac{1}{2}\pb_{\psi}\right)^m\left[\lambda+(-1+2\delta_{m,1})\lambda^{(0)} \right]
\ee
where $[m]$ denotes the order in $\psi$, as before. At this point, we recall again that both $\lambda$ and $\psi$ are linear in $x^i$ (\eqref{ham_def} and \eqref{psi_n_lim_xi}), which, in conjunction with the definition of the Poisson bracket $\pb$ (\eqref{pb_notation} and \eqref{poisson_def}) implies that only the $m=0$ and $m=1$ terms will contribute, so we have the remarkably simple expression
\be\label{delta_psi_all_ords}
\delta_\lambda \psi=\lambda-\lambda^{(0)} 
+ \frac{1}{4}\pb_{\psi}\left[\lambda+\lambda^{(0)} \right]
\ee
In \autoref{Appendix_consistency_check_grav_full} we show that this satisfies the consistency condition \eqref{consistency_cond_grav_full}, as needed, thus completing the proof for the double copy relations.

\section{Extension in Bondi coordinates} \label{extension_Bondi_coord}

We will study fields near future null infinity, $\Ib^+$, so we are switching to coordinates adapted to it. The natural choice are Bondi-type coordinates $(r,u,z, \bar{z})$  given by
\be \label{Bondi_coordinates}
r = V , \quad z = \frac{Z}{V} , \quad \bar{z} = \frac{\Zb}{V}, \quad u = U - \frac{Z \Zb}{V}.
\ee
In these coordinates, the Minkowski metric reads
\be 
ds^2 = -2dudr + 2 r^2dz d\bar{z},
\ee
where we see that by taking the conformally rescaled metric $\frac{1}{r^2} ds^2$, we have a well defined (degenerate) metric on $\Ib^+$ given by $2dz d\bar{z}$.\footnote{See e.g. Section 4 of \cite{Barnich:2010eb} for a derivation of this metric from general assumptions about asymptotic flatness. We remark that these coordinates were chosen due to the simplicity of the map to the light-cone coordinates. One could also perform the analysis in standard Bondi coordinates, resulting in slightly more complicated expressions.}

The radiative data of a massless scalar field is given by 
\be \label{phi_fall_off_scri}
\varphi(r,u,z,\bar{z}) = \frac{\varphi_\Ib(u,z, \bar{z})}{r} + \mathcal{O}(r^{-2})
\ee
as $r \rightarrow +\infty$.

\subsection{Phase space for SDYM and fields near infinity}
 Let us use the notation
\be
\A_z(r,u,z,\zb)  \stackrel{r \to \infty}{=}  A_z(u,z,\zb) + \mathcal{O}(r^{-1}) \label{Afallr}
\ee
\black Then, considering equations \eqref{Aphi} in Bondi coordinates \eqref{Bondi_coordinates}, we have
\be \label{A_z_Bondi}
\A_z = r \partial_u \Phi , \quad \A_{\bar{z}} = 0,
\ee
The \textit{radiative phase space} will then be given by $\Phi_\Ib$ as free data,
\be 
\Gamma^{rad} = \{ \Phi_\Ib (u,z,\bar{z}) \mid A_z= \partial_u \Phi_{\I}, \   A_{\zb}  = 0  \}.
\ee
where we have used \eqref{phi_fall_off_scri}-\eqref{A_z_Bondi}.

Recall that the gauge parameters depend only on $y$ coordinates, $\Lambda(y)=\Lambda(V,Z)$, which in this case are $(r,z)$. Therefore, we can assume 
\be \label{Lambda_r_exp}
\Lambda = \sum_{n = -\infty }^{+\infty} \Lambda^{(n)} r^n,
\ee
as an $r-$expansion for the gauge parameter, with $ \Lambda^{(n)}= \Lambda^{(n)}(z)$. The first residual gauge symmetry is given by $O(r^0)$, 
\be 
\Lambda = \Lambda^{(0)}(z) + ...
\ee
The induced transformation on the phase space,
\be 
\delta_{\Lambda} \Phi_\Ib = u \partial_z \Lambda^{(0)}+ i [\Lambda^{(0)}, \Phi_\Ib]
\ee
induces the leading asymptotic symmetry. The extended phase space is now:  
\be \label{ext_space_full_YM_asympt}
\Gamma^{\text{ext}}_{\infty,\text{YM}}:=\Gamma^{rad}_{YM} \times \{\Psi \mid \Psi = \sum_{n=1}^{+\infty} r^n \Psi^{(n)} (z) \}.
\ee 
with the extended gauge field constructed as in \eqref{A_tilde_full}, with the approprite coordinate change. The transformation of $\Psi$ to all orders (equation \eqref{solved_transform_all_orders}) then immediately translates to our Bondi coordinates
\be
\left[\delta^{[m]}_\Lambda \Psi\right]^{(n)}=\frac{B_m^+}{m!}\left\{\left(ad_{-i\Psi}\right)^m\left[\Lambda+(-1+2\delta_{m,1})\Lambda^{(0)} \right]\right\}^{(n)}
\ee
with $\Lambda$ as in \eqref{Lambda_r_exp}, $\Psi$ as in \eqref{ext_space_full_YM_asympt}, $[m]$ denoting the order in field fluctuations as before, and $(n)$ now denoting the order in $r$.

We expect these LGTs to be associated with sub$^{(n)}$-leading charges \cite{Hamada:2018vrw,Campiglia:2018dyi}, whose construction we leave for future work. The linear extension was already done in \cite{Campiglia:2021oqz}, while it remains to be shown in general. In this work, we propose a first step towards this general result, in particular we look at how the variation algebra can be computed when considering SDYM.  

Take two large gauge transformations with parameters,
\be 
\Lambda_1 = \sum_{n = -\infty }^{N_1} \Lambda_1^{(n)} r^n, \quad \Lambda_2 = \sum_{n = -\infty }^{N_2} \Lambda_2^{(n)} r^n,
\ee
then 
\be 
[\Lambda_1 , \Lambda_2] = r^{N_1 + N_2 } [\Lambda_1^{(N_1)} , \Lambda_2^{(N_2)}] + O(r^{N_1 + N_2 -1}).
\ee
 Then, truncating $\delta_\lambda\Psi$ up to order $O(\Psi^{(N_1 + N_2)})$, one can verify 
\be 
[\overset{N_1+N_2}{\delta}_{\Lambda_1} , \overset{N_1+N_2}{\delta}_{\Lambda_2}] \Psi =- i\overset{N_1+N_2}{\delta}_{[\Lambda_1 , \Lambda_2]} \Psi.
\ee
thus showing that this is a consistent truncation.

\subsection{Phase space for SDG}
For gravity, we will use the notation
\be
h_{zz}(r,u,z,\zb)  \stackrel{r \to \infty}{=}  r C_{zz}(u,z,\zb) +  \mathcal{O}(r^0) \label{hfallr}
\ee
\black
where $C_{z z}$ is the \textit{shear tensor}.
The scalar potential $\phi_\Ib$ defines a decay for $h_{zz}$ as follows, 
\be 
h_{zz} (r,u,z,\bar{z}) = r \partial^2_u \phi_\Ib + \mathcal{O}(r),\quad h_{\zb \zb}=0
\ee
The phase space is defined analogously with the YM case
\be 
\Gamma^{rad} = \{ \phi_\Ib (u,z,\bar{z})\mid  C_{zz} = \partial^2_u \phi_{\I}, \   C_{\zb \zb}  = 0  \}.
\ee
where we used \eqref{phi_fall_off_scri} and \eqref{hfallr}. The action of the residual symmetry \eqref{xiizero} on the radiative phase space corresponds to the fall-off 
\be
b(r,z) =-r f(z). \label{bitof}
\ee
and then
\be
\delta_b \phi_{\I} = - u^2  f''(z)  + f(z) \partial_u \phi_{\I} \label{delfphi}
\ee
corresponding to a holomorphic supertranslation with parameter $f(z)$. To allow the Hamiltonian $\lambda$ given in \eqref{ham_def} to go to to all orders in r, we introduce the extended phase space as in \eqref{ext_space_full_gravity}, resulting in the transformation rules \eqref{delta_psi_all_ords}, translated into Bondi coordinates. 

For the family of LGT that we are taking, \eqref{xiizero}, we can construct directly the variation algebra from the double copy prescription given in the last section. In terms of the charges, it is not clear whether sub$^{n}$-leading soft charges correspond to higher order LGT, although it was conjectured that certain $O(r)$ diffeomorphisms could be associated to the charge corresponding to the sub-sub-leading soft graviton theorem, \cite{Campiglia:2016efb}. It would be interesting to explore further this direction. 

Finally we note that the transformations discussed in this section correspond to ($\mathcal{O}(r^n)$ extensions of) supertranslations. It turns out that the self-dual sector also accommodates superrotations within its rich collection of residual symmetries. This will be further detailed in the next section. 

\section{Discussion: An infinite cube of symmetries} \label{An infinite cube of symmetries in the self-dual sector} 
Self-dual YM and gravity, being integrable, have an in finite number of symmetries \cite{Popov:1996uu,Popov:1998pc,Dolan:1983bp,Prasad:1979zc,Park:1989vq,Husain:1993dp,Husain:1994vi}. These can be defined recursively as follows. Assume $\delta_1\Phi$ is a symmetry of the SDYM equation \eqref{eomPhi}, i.e.
\be
\square \delta_1 \Phi  =  - 2 i \Pi^{i j}[\partial_i \Phi,\partial_j \delta_1 \Phi]  \label{eomdelPhi} 
\ee
 Then we can define a new symmetry $\delta_2\Phi$ via
\be
\partial_i \delta_2 \Phi =  \Omega_i^{\ \alpha} \partial_\alpha \delta_1 \Phi - i [\partial_i \Phi, \delta_1 \Phi] \label{defdeltPhi} 
\ee
Thus we can start with $\delta_1\Phi$ as defined in \eqref{delA} and \eqref{lam} and proceed indefinitely to construct an infinite tower of symmetries. The method works identically for gravity, and the double copy relations continue to hold at every order \cite{Campiglia:2021srh}:
\be\label{diag}
\begin{array}{ccccc}
\text{\textbf{gauge}}\qquad& \delta_{1} \Phi & \xrightarrow{\text{DC}} & \delta_{1} \phi  &\qquad\text{\textbf{diffeo}}_1  \\
&  \downarrow & & \downarrow &\\
\text{\textit{extra}}\qquad&  \delta_{2} \Phi  & \xrightarrow{\text{DC}} & \delta_{2} \phi &\qquad \text{\textbf{diffeo}}_2\\
&  \downarrow & & \downarrow &\\
&  \vdots & & \vdots & \\
\text{\textit{extra}}\qquad&  \delta_{n} \Phi  & \xrightarrow{\text{DC}} & \delta_{n} \phi &\qquad\text{\textit{extra}}\\
&  \vdots & & \vdots &
\end{array}
\ee
Note that $\delta_1\Phi$ and $\delta_1\phi$ are subsets of the standard gauge and diffeomorphism transformations respectively. $\delta_2\phi$ turns out to be a different subset of diffeomorphisms \cite{Campiglia:2021srh} \footnote{Specifically, the transformation parameter is given by
\be
\xi_i = - c_i(y) , \quad \xi_\alpha =   \partial_\alpha c_j(y) x^j . \label{xic}
\ee
}. Intriguingly, $\delta_2\Phi$, its YM origin in the double copy, \textit{is not a subset of gauge transformations}. This might suggest that a broader framework is needed in order to recover gravitational symmetries via the double copy. Conversely, there is an interesting programme aiming to reveal the existence of a wider (perhaps infinite) set of symmetries in gravity and YM \cite{Freidel:2021ytz}. In this context, one might speculate that the \textit{extra} symmetries in \eqref{diag} could be revealed as genuine symmetries of the full (i.e. non-self-dual) theories.

\begin{figure}[H]
\hspace*{30pt}
\includegraphics[scale=0.5]{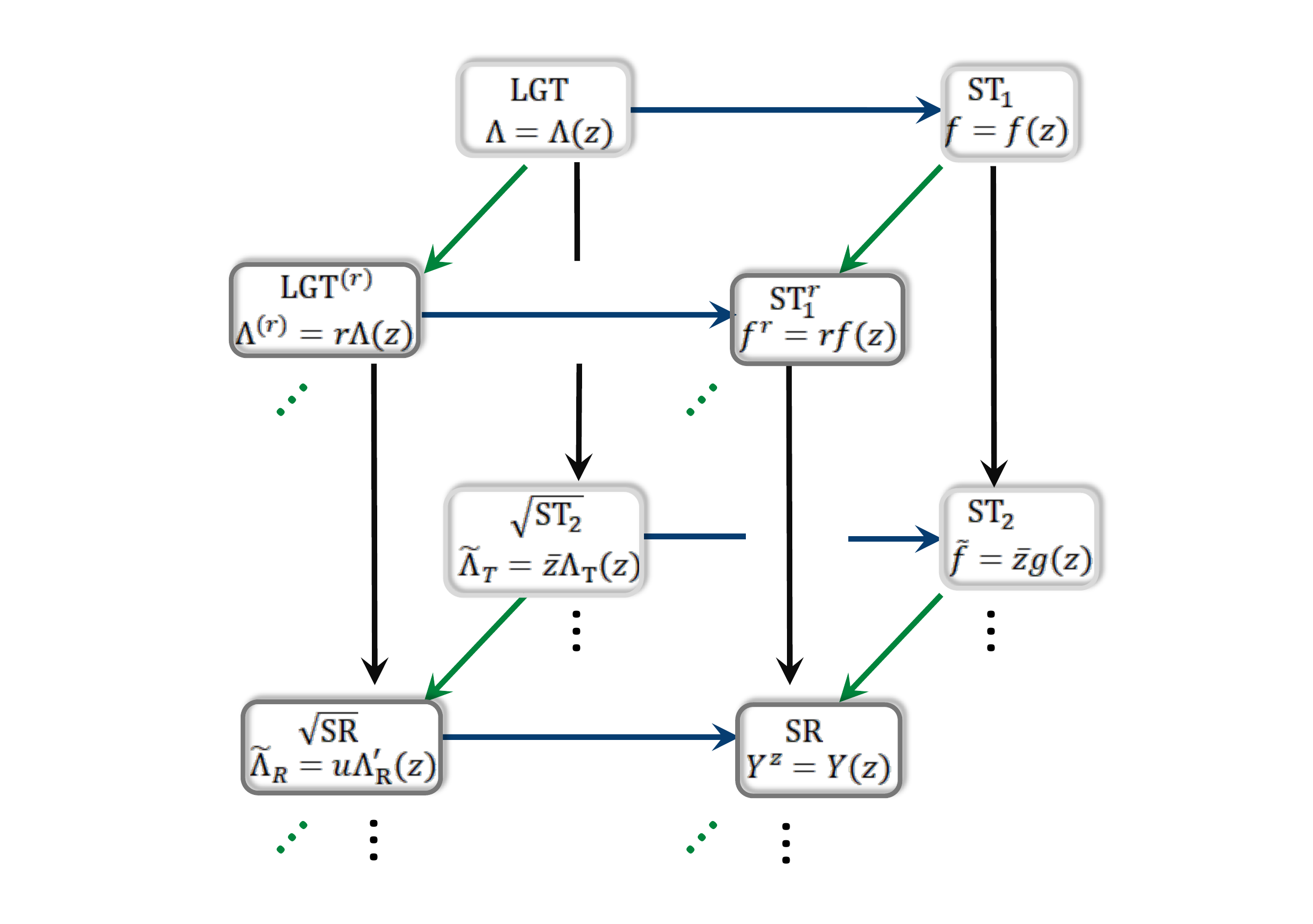}
\caption{}
\label{fig:asymptsumm_new} 
\end{figure}

How does this relate to the subject of the present article? To see this, we first go to null infinity, working in flat Bondi coordinates described in the previous section, to summarise the results of \cite{Campiglia:2021srh} in the picture below: 

\begin{figure}[H]
\centering
\includegraphics[scale=0.5]{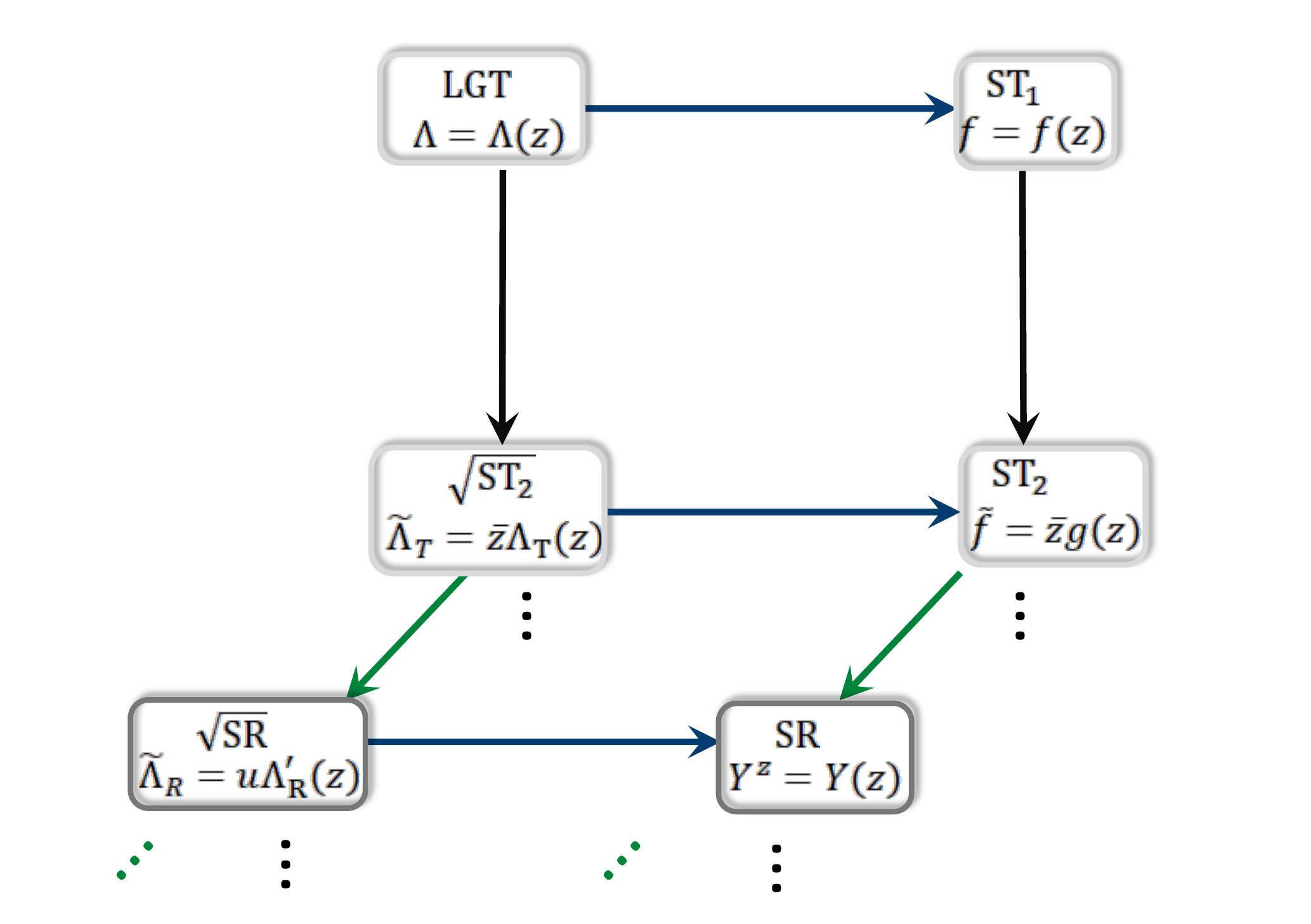}
\caption{}
\label{fig:asymptsumm} 
\end{figure}
In the above, LGT stands for non-abelian large gauge transformations, ST and SR for supertranslations and superrotations.  Horizontal lines represent the double copy relations. Vertical lines represent the symmetry raising map \eqref{defdeltPhi}  between different families of symmetries.  The diagonal arrows connecting different symmetries in the 2nd family denote an increase in the power of $r$ in the "seed" symmetries belonging to family 1 that need to be plugged into \eqref{defdeltPhi} in order to generate them. For example the $\sqrt{SR}$ symmetry with parameter $u\Lambda_r'(z)$ was obtained by plugging a  family 1 gauge symmetry with parameter $r\Lambda(z)$ into the recursion relation \eqref{defdeltPhi} (see \cite{Campiglia:2021srh}). However, the transformation with parameter $r\Lambda(z)$ was not actually allowed, as it violated the fall-off! The situation is rectified in the current work, via the definition of the extended phase space, which reinterprets the violating transformation as a legitimate symmetry of the extended phase space, acting on $\Psi$ (and analogously for gravity). This means that we are now allowed to add diagonal rungs to the figure above, corresponding to increasing powers of $r$, as they correspond to legitimate symmetries in our interpretation - see \autoref{fig:asymptsumm_new}. This completes the construction of the various relations between the infinite tower of symmetries in SDYM and gravity. We summarise these in \autoref{fig:asymptsumm_new} below, which is understood to extend infinitely in the vertical and diagonal directions. 

\black

\section{Conclusions} \label{Conclusions}

In this paper we showed how to construct an extended phase space to all orders in the expansion coordinate. This is important for studying the subleading (to all orders) behaviour of soft gluons and gravitons. Our analysis was restricted to the self-dual sectors of gravity and YM. This allowed us to additionally establish a double copy map for a subset of finite local transformations in these sectors. 

We recall that the self-dual sectors of gravity and YM are integrable, thus possessing infinite towers of symmetries  \cite{Popov:1996uu,Popov:1998pc,Dolan:1983bp,Prasad:1979zc,Parkes:1992rz,Husain1994,Husain1995,Wolf2005,Popov2007}, defined recursively, as described in the previous section (see \eqref{diag}). In the present work, we have restricted ourselves to only the first family of symmetries, however to all orders in $r$. The various interconnections between the symmetries are sketched in \autoref{An infinite cube of symmetries in the self-dual sector}  (see also \autoref{fig:asymptsumm_new} above). In would be interesting to connect with the celestial holography programme, where the sub$^n$-leading effects in amplitudes are naturally incorporated, and a similarly rich algebraic structure emerges \cite{Cheung:2016iub,Nande:2017dba,He:2015zea,Himwich:2019dug,Fan:2019emx,Kalyanapuram:2021tnl,Raclariu:2021zjz,Pasterski:2021rjz,Guevara:2021abz,Jiang:2021ovh,Schwarz:2022dqf,Pasterski:2022jzc,Albayrak:2020saa,Gonzalez:2020tpi,Gonzalez:2021dxw,Magnea:2021fvy}.

On the other hand, we are looking at extending our procedure for the all-orders phase space beyond the self-dual sectors.\footnote{See \cite{Ashtekar2021} for a novel way to connect between the self-dual sector and the full theory, in the Hamiltonian formalism.}  Here a natural starting point will be to work in the light-cone gauge. Even for the full theories, this is more restrictive than the Lorenz and De Donder gauge respectively - this gives a simpler set of residual symmetries at null infinity and is thus a good starting point for constructing extended phase spaces to all orders. 

\acknowledgments
The authors are grateful to Miguel Campiglia for valuable discussions. SN was supported by an SFI-IRC Pathway grant 21/PATH-S/9391. JP is partially supported by CAP PhD scholarship, by CSIC grant C013-347. JP also wants to thank the hospitality of DIAS (Dublin, Ireland) during a visit in 2022 related to the work on this paper.

\appendix

\section{YM calculations}\label{Full YM transformation}

\subsection{Linear extended phase space transformation}\label{Some_YM_calcs}

Consider \eqref{cons_lin_YM_simplified}, reproduced below for convenience
\be \label{gen_lin_constr}
D_\alpha \delta_{\Lambda} \Psi = -i[D_\alpha \Lambda_{+} ,\Psi] + D_\alpha \left(\Lambda_{+ } -i [\Psi, \Lambda] \right).
\ee
At order $V^0$, for $\alpha = Z$,
\beq 
\partial_Z (\delta_{\Lambda} \Psi)^{(0)} -i \sum_{n=0}^{\infty} [A_Z^{(-n)}, (\delta_{\Lambda} \Psi)^{(n)}] &=& -i([\partial_Z \Lambda_{+} ,\Psi])^{(0)} - \sum_{k=0}^{\infty} [\sum_{n=0}^{\infty}[A_Z^{(-n)}, \Lambda_{+}^{(n-k)}] ,\Psi^{(k)}]\nonumber \\
&&+ \partial_Z \left(\Lambda_{+ } -i [\Psi, \Lambda] \right)^{(0)} -i \sum_{n=0}^{\infty} [A_Z^{(-n)}, \Lambda_{+ }^{(n)}]\nonumber \\
&& -\sum_{k=0}^{\infty} \sum_{n=0}^{\infty} [A_Z^{(-n)},  [\Psi^{(k)}, \Lambda^{ n-k}]]
\eeq
Note that several terms in the above vanish at $\mathcal{O}(V^0)$, specifically those involving only $\Psi$ and $\Lambda_+$, due to \eqref{ext_space_lin_YM} and \eqref{splitting_Lambda}. Then,
\beq 
-i \sum_{n=1}^{\infty} [A_Z^{(-n)}, (\delta_{\Lambda} \Psi)^{(n)}] &=& \sum_{k=1}^{\infty} \sum_{n=1}^{\infty}
 [A_Z^{(-n-k)}, \Lambda_{+}^{(n)}] ,\Psi^{(k)}]\nonumber \\
&& -i \partial_Z \left([\Psi, \Lambda_-] \right)^{(0)}  -i \sum_{n=1}^{\infty} [A_Z^{(-n)}, \Lambda_{+ }^{(n)}]  \nonumber\\
&& -\sum_{k=1}^{\infty} \sum_{n=0}^{\infty} [A_Z^{(-n)},  [\Psi^{(k)}, \Lambda^{( n-k)}]]
\eeq
Remembering that $\Psi^{(n)}$ and $\Lambda_+^{(n)}$ were arbitrary we can, in a first instance, choose $\Psi^{(n)}=0,\ n>1$, and same for $\Lambda_+$. We will also neglect small gauge transformations, thus only keeping $\Lambda_-^{0}$ and $\Lambda_+^{1}$. Then
\beq 
-i [A_Z^{(-1)}, (\delta_{\Lambda} \Psi)^{(1)}] &=& [[ A_Z^{(-2)}, \Lambda_{+}^{(1)}] ,\Psi^{(1)}]  -i [A_Z^{(-1)}, \Lambda_{+ }^{(1)}]  \nonumber \\
&&-[A_Z^{(-2)},  [\Psi^{(1)}, \Lambda_+^{(1)}]] -[A_Z^{(-1)},  [\Psi^{(1)}, \Lambda_-^{(0)}]] 
\eeq
At this stage, as in \cite{Campiglia:2021oqz}, we are forced to employ an additional simplification, and require that we only work to linear order in the new fields and tranformations, i.e. $\Psi$ and $\Lambda_+$. The above then reduces to 
\be
-i [A_Z^{(-1)}, (\delta_{\Lambda} \Psi)^{(1)}]=-i [A_Z^{(-1)}, \Lambda_{+}^{(1)}-i[\Psi^{(1)}, \Lambda_-^{(0)}]]
\ee
finally allowing us to read off
\be \label{delta_psi_1_linear}
(\delta_{\Lambda} \Psi)^{(1)} =\Lambda_{+}^{(1)}-i[\Psi^{(1)}, \Lambda_-^{(0)}]
\ee
As a sanity check, we can also attempt to look at order $V^1$ in \eqref{gen_lin_constr}. With the restrictions above, this reduces to:
\be 
D_Z^{(0)} \delta_{\Lambda} \Psi^{(1)} = D_Z^{(0)} \left(\Lambda_{+ } -i [\Psi, \Lambda] \right)^{(1)}.
\ee
which is in agreement with \eqref{delta_psi_1_linear}. 

Now let us attempt to extend these results to second order in $V$, i.e. consider $\Psi^{1}$, $\Psi^{2}$ and $\Lambda_-^{0}$, $\Lambda_+^{1}$, $\Lambda_+^{2}$. In this case,  \eqref{gen_lin_constr} should hold at order $V^2$. We get
\be \label{ugly_o_2}
 D_Z^{(0)} \delta_\Lambda \Psi^{(2)}= -i[D^{(0)}_Z \Lambda^{(1)} , \Psi^{(1)}] + D^{(0)}_Z \left( \Lambda^{(2)} -i [ \Psi^{(1)},\Lambda^{(1)}] -i [\Psi^{(2)},\Lambda^{(0)} ] \right)
\ee
If we apply the linearization with respect to LGT's and $\psi$, again the terms $[\Psi, \Lambda_+]$ are thrown away, obtaining
\be \label{delta_psi_2_linear}
(\delta_{\Lambda} \Psi)^{(2)} =\Lambda_{+}^{(2)}-i[\Psi^{(2)}, \Lambda_-^{(0)}].
\ee
Nevertheless, we will show that is not the desirable result when considering higher order LGTs. The linearisation w.r.t. $\Psi$ should be preserved by the algebra, i.e. we want the variations to have the correct representation on the phase space functions\footnote{Recall that $\overset{n}{\delta}$ denotes the variation up to order $n$ in $\Psi$.}, 
\be 
[\overset{1}{\delta}_{\Lambda_1} , \overset{1}{\delta}_{\Lambda_2}]\Psi  =- i\overset{1}{\delta}_{[\Lambda_1, \Lambda_2]} \Psi.
\ee
Take two LGTs, let say $\Lambda_1$ and $\Lambda_2$. For simplicity, let us set $\Lambda_{+}^{(n)}=0$, $n\geq2$ for both of them.  Then, 
\be 
[\Lambda_1 , \Lambda_2] = V^2[\Lambda_1^{(1)}, \Lambda_2^{(1)}] + V([\Lambda_1^{(1)}, \Lambda_2^{(0)}] + [\Lambda_1^{(0)}, \Lambda_2^{(1)}]) + [\Lambda_0^{(0)}, \Lambda_2^{(1)}] ,
\ee
Our linearisation assumptions would now dictate that we should throw away the first term in the equation above. But this would now not reflect the non-abelian nature of the YM theory. The problem will persist for higher order LGTs, i.e. $[\Lambda_1^{(n)}, \Lambda_2^{(m)}]=0$ for all $n,m>0$.


 For $[\Lambda_1^{(1)}, \Lambda_2^{(1)}]$ to appear we now have to relax our linearisation assumptions,  in other words we are seeking a solution to \eqref{ugly_o_2} up to second order in  LGT's and $\psi$. It is straightforward to see that this does not admit a field-independent, local solution.

\subsection{Equation \eqref{rearranged_YM_full_cond} }\label{sandwich}
We start with \eqref{consistency_full_YM}, reproduced below for convenience 
\be \label{4.3_copy}
\delta_\Lambda \hat{\Av}_\alpha = \hat{D}_\alpha \Lambda
\ee
The RHS can be rearranged as
\be\label{4.3_copy_RHS}
\begin{aligned}
\hat{D}_\alpha \Lambda&=\partial_\alpha\Lambda-i[\hat{\Av}_\alpha ,\Lambda]\\
&=\partial_\alpha\Lambda-i[e^{i\Psi} \Av_\alpha e^{-i\Psi} + ie^{i\Psi} \partial_\alpha e^{-i\Psi} ,\Lambda]\\
&=\partial_\alpha\Lambda+[e^{i\Psi} \partial_\alpha e^{-i\Psi} ,\Lambda]-i[e^{i\Psi} \Av_\alpha e^{-i\Psi},\Lambda]\\ 
&=e^{i\Psi}\partial_\alpha\left(e^{-i\Psi}\Lambda e^{i\Psi} \right)e^{-i\Psi}
-i e^{i\Psi} [\Av_\alpha,e^{-i\Psi}\Lambda e^{i\Psi}]e^{-i\Psi}\\
&=e^{i\Psi}D_\alpha\left(e^{-i\Psi}\Lambda e^{i\Psi} \right)e^{-i\Psi}
\end{aligned} 
\ee
and the LHS:
\be\label{4.3_copy_LHS}
\begin{aligned}
\delta_\Lambda \hat{\Av}_\alpha =&\delta_\Lambda (e^{i\Psi} \Av_\alpha e^{-i\Psi} + ie^{i\Psi} \partial_\alpha e^{-i\Psi})\\
=&-\Op_{-i\Psi}(-i\delta_\Lambda\Psi)e^{i\Psi}\Av_\alpha e^{-i\Psi} + e^{i\Psi} (\delta_\Lambda\Av_\alpha) e^{-i\Psi}+ e^{i\Psi} \Av_\alpha e^{-i\Psi}\Op_{-i\Psi}(-i\delta_\Lambda\Psi)\\
&-i\Op_{-i\Psi}(-i\delta_\Lambda\Psi)e^{i\Psi} \partial_\alpha e^{-i\Psi}+ ie^{i\Psi} \partial_\alpha\left(e^{-i\Psi}\Op_{-i\Psi}(-i\delta_\Lambda\Psi) \right)\\
&=e^{i\Psi}\Big\{\delta_\Lambda\Av_\alpha +\Av_\alpha e^{-i\Psi}\Op_{-i\Psi}(-i\delta_\Lambda\Psi)e^{i\Psi}-e^{-i\Psi}\Op_{-i\Psi}(-i\delta_\Lambda\Psi)e^{i\Psi}\Av_\alpha\\
&\qquad\quad +ie^{-i\Psi}\Op_{-i\Psi}(-i\delta_\Lambda\Psi)\partial_\alpha e^{i\Psi} 
+i\partial_\alpha\left(e^{-i\Psi}\Op_{-i\Psi}(-i\delta_\Lambda\Psi)\right)e^{i\Psi} \Big\} e^{-i\Psi}\\
=&e^{i\Psi}\Big\{\delta_\Lambda\Av_\alpha -i\left[\Av_\alpha,e^{-i\Psi}\Op_{-i\Psi}(\delta_\Lambda\Psi)e^{i\Psi} \right]
+\partial_\alpha\left(e^{-i\Psi}\Op_{-i\Psi}(\delta_\Lambda\Psi)e^{i\Psi} \right)\Big\} e^{-i\Psi}\\
=&e^{i\Psi}\Big[\delta_\Lambda\Av_\alpha
+D_\alpha\left(e^{-i\Psi}\Op_{-i\Psi}(\delta_\Lambda\Psi)e^{i\Psi} \right)\Big] e^{-i\Psi}
\end{aligned} 
\ee
where we used
\be
\delta e^X=e^X \Op_X (\delta X)\quad \Rightarrow \quad \delta e^{-X}=-\Op_X (\delta X)e^{-X}
\ee
with $\Op_X$ as defined in \eqref{O_operator_def} and we set $X=-i\Psi$. Finally, equation \eqref{rearranged_YM_full_cond} follows from \eqref{4.3_copy}, \eqref{4.3_copy_LHS} and \eqref{4.3_copy_RHS}.

\subsection{Extended phase space transformation at all orders} \label{Extended phase space transformation at all orders}
We wish to find an explicit perturbative expansion of equation \eqref{delta_Psi_formal}, reproduced below for convenience, working at some arbitrary order $m$ in $\psi$:
\be \label{formal_inv_eqn_for_delta}
\delta^{[m]}_\Lambda \Psi = \left( \Op^{-1}_{-i\Psi}(\Lambda - e^{i\Psi}\Lambda^{(0)}e^{-i\Psi}) \right)^{[m]}.
\ee
with 
\be
\Op_{-i\Psi} :=  \frac{1 - e^{-ad_{-i\Psi}}}{ad_{-i\Psi}}
\ee
Using
\be \label{exp_identity_comm}
e^B A e^{-B}=e^{ad_B}A \ , 
\ee
we can rewrite \eqref{formal_inv_eqn_for_delta} as
\be\label{rewritten formal}
\delta^{[m]}\delta_\Lambda \Psi = \left( \Op^{-1}_{-i\Psi}\Lambda - \Op^{-1}_{-i\Psi}e^{ad_{i\Psi}}\Lambda^{(0)} \right)^{[m]} 
\ee
At this point, to simplify calculations, we introduce the notation
\be\label{Y_def}
Y\equiv ad_{-i\Psi} 
\ee
Then we can perturbatively invert the operator
\be 
\Op_{-i\Psi}=\frac{1-e^{-Y}}{Y}\equiv Y^{-1} (1-e^{-Y})
\ee
by making use of the expansion
\be
(1-e^{-Y})^{-1}Y\equiv\frac{Y}{1-e^{-Y}}=\sum_{m=0}^{\infty}\frac{B_m^+ Y^m}{m!}
\ee
where $B^+_n$ are the Bernoulli numbers $B^+_0=1,\ B^+_1=\frac{1}{2},\ B^+_2=\frac{1}{6},\ ...$ 

The first term in \eqref{rewritten formal} is then simply
\be \label{1st_ptC17}
 \left( \Op^{-1}_{-i\Psi}\Lambda\right)^{[m]} =\frac{B_m^+}{m!}Y^m(\Lambda)
\ee
The second term requres a little more work. We have
\be
\begin{aligned}
- \left(\Op^{-1}_{-i\Psi}e^{-Y}\Lambda^{(0)} \right)^{[m]}
&=-\sum_{k=0}^{m}\left(\Op^{-1}_{-i\Psi}\right)^{[k]} \left( e^{-Y}\Lambda^{(0)} \right)^{[m-k]}\\
&=-\sum_{k=0}^{m}\frac{B_k^+ Y^k}{k!}(-1)^{m-k}\frac{Y^{m-k}}{(m-k)!}\Lambda^{(0)}\\
&=-\sum_{k=0}^{m}(-1)^{m-k}B_k^+\frac{1}{k!(m-k)!}Y^m\Lambda^{(0)}
\end{aligned} 
\ee
Now we use $(-1)^kB_k^+=B_k^-$ (this follows from the fact that all odd Bernoulli numbers vanish, except $B_1^-=-B_1^+$) to write
\be
\begin{aligned}
- \left(\Op^{-1}_{-i\Psi}e^{-Y}\Lambda^{(0)} \right)^{[m]}
&=-(-1)^m \sum_{k=0}^{m} B_k^- \frac{1}{k!(m-k)!}Y^m\Lambda^{(0)}\\
&=\left[-(-1)^m\frac{B_m^-}{m!}-(-1)^m \sum_{k=0}^{m-1}  B_k^- \frac{1}{k!(m-k)!}\right]Y^m\Lambda^{(0)}\\
&=\left[ -\frac{B_m^+}{m!}-\frac{(-1)^m}{m!} \sum_{k=0}^{m-1}  B_k^- \binom{m}{k} \right]Y^m\Lambda^{(0)}
\end{aligned} 
\ee
Using the sum formula for the Bernoulli numbers
\be 
\sum_{k=0}^n\binom{n+1}{k}B_k^-=\delta_{n,0}
\ee
the equation above finally becomes
\be \label{2nd_ptC17}
\begin{aligned}
- \left(\Op^{-1}_{-i\Psi}e^{-Y}\Lambda^{(0)} \right)^{[m]}
&=\left[ -\frac{B_m^+}{m!}-\frac{(-1)^m}{m!} \delta_{m-1,0} \right]Y^m\Lambda^{(0)}\\
&=\left[ -\frac{B_m^+}{m!}+\delta_{m,1} \right]Y^m\Lambda^{(0)}\\
&=\left[ -\frac{B_m^+}{m!}+\frac{2B_m^+}{m!}\delta_{m,1}\right]Y^m\Lambda^{(0)}\\
&=\frac{B_m^+}{m!}(-1+2\delta_{m,1})Y^m\Lambda^{(0)}
\end{aligned} 
\ee
where to get to the third line we used $B_1^+=\tfrac{1}{2}$. Finally, putting together \eqref{1st_ptC17} and \eqref{2nd_ptC17}, and using \eqref{Y_def}, we get
\be 
\delta^{[m]}_\Lambda \Psi=\frac{B_m^+}{m!}\left(ad_{-i\Psi}\right)^m\left[\Lambda+(-1+2\delta_{m,1})\Lambda^{(0)} \right]
\ee

\section{Gravity Calculations}

\subsection{Restricted diffeomorphism transformation}\label{Restricted diffeomorphism transformation}
As explained in \autoref{Light-cone gauge in the self-dual sectors of YM and gravity}, we are working with a subset of diffeomorphisms parametrised by $\xi_{i}=0$ and $\xi_\alpha=b_\alpha(y)$. The transformation of the metric $h_{\alpha \beta} = \Pi_\alpha^{\ i} \Pi_\beta^{\ j} \partial_i \partial_j \phi \label{hphi}$ can then be written in terms of the "Hamiltonian" $\lambda = 2 \Omega_i^{\ \alpha} x^i b_\alpha$ as 
\be
\begin{aligned}
\delta h_{\alpha\beta}&=\L_\xi g_{\alpha \beta}=\L_\xi \eta_{\alpha \beta} + \L_\xi h_{\alpha \beta} \\
&=\Pi_{(\alpha}^{\ i}\partial_i \left( \partial_{\beta)} \lambda - \tfrac{1}{2}\Pi_{\beta)}^{\ j}\partial_j\{\lambda,\phi \}\right)\\
&=\Pi_{(\alpha}^i \partial_i \partial_{\beta )} \lambda - \frac{1}{2} \left(   \{ \Pi_{(\alpha}^i \partial_i  \Pi_{\beta)}^j \partial_j \lambda , \phi \} + \{ \lambda , \Pi_{(\alpha}^i \partial_i  \Pi_{\beta)}^j \partial_j \phi\} + 2 \{ \Pi_{(\alpha}^i \partial_i \lambda , \Pi_{\beta)}^j \partial_j \phi \} \right)\\
&=\Pi_{(\alpha}^i \partial_i \partial_{\beta )} \lambda - \frac{1}{2} \{ \lambda ,h_{\alpha \beta} \},
\end{aligned} 
\ee
where we used the fact that $\lambda$ is at most linear in $x^i$, together with the definition of the Poisson bracket \eqref{poisson_def}.

\subsection{Poisson bracket exponentiation}\label{Full gravity calculations}
Start with (see \autoref{Restricted diffeomorphism transformation})
\be
 \Li_\xi g_{\alpha \beta} = \Pi_{(\alpha}^i \partial_i \partial_{\beta )} \lambda - \frac{1}{2} \{ \lambda ,h_{\alpha \beta} \}.
\ee
Next, we compute:
\beq
\Li_\xi (\Li_\xi ( g_{\alpha \beta})) &=& \Li_\xi ( \Pi_{(\alpha}^i \partial_i \partial_{\beta )} \lambda - \frac{1}{2} \{ \lambda ,h_{\alpha \beta} \}) \\  \label{D3_eqn}
&=& \xi^k \partial_k \left(\Pi_{(\alpha}^i \partial_i \partial_{\beta )} \lambda - \frac{1}{2} \{ \lambda ,h_{\alpha \beta} \} \right) \\ \label{D4_eqn}
&=& -\frac{1}{2} \xi^k \partial_k \{ \lambda , h_{\alpha \beta}\}\\ \label{D5_eqn}
&=& \frac{1}{4} \{\lambda , \{ \lambda , h_{\alpha \beta}\}\}.
\eeq
In the above, to get \eqref{D3_eqn}, we used the definition of the Lie derivative, together with our restricted diffeomorphism parameter \eqref{xiizero} and the fact that $(\Li_\xi g)_{\alpha i}=0$ by construction. Then, \eqref{D4_eqn} folows if we recall that $\lambda$ is at most linear in $x^i$ (see \eqref{ham_def}). Finally, we get \eqref{D5_eqn} by using \eqref{useful_poisson_xi}.

We can immediately see that for $n>1$,
\be \label{D6_eqn}
(\Li_\xi) ^{n} g_{\alpha \beta} = \frac{1}{(-2)^n} \pb^n_\lambda (h_{\alpha \beta})\ ,
\ee
which follows by the considerations above and repeated use of \eqref{useful_poisson_xi}.

\subsection{Consistency check}\label{Appendix_consistency_check_grav_full}
We wish to check the consistency condition \eqref{consistency_cond_grav_full}, copied below for convenience
\be
\delta \hat{h}_{\alpha\beta}= \Pi_{(\alpha}^i \partial_i \partial_{\beta )} \lambda - \frac{1}{2}  \pb_\lambda \hat{h}_{\alpha \beta} \ ,
\ee
making use of \eqref{hat_h_with_exp}, together with the transformation rule for $h_{\alpha\beta}$, which is unchanged by the extension of $\lambda$ to higher orders in $V$:
\be 
\delta h_{\alpha\beta}=\Pi_{(\alpha}^i \partial_i \partial_{\beta )} \lambda^{(0)} - \frac{1}{2} \pb_{\lambda^{(0)}} h_{\alpha \beta},
\ee
and the transfomation rule for $\psi$ at all orders in $V$ \eqref{delta_psi_all_ords}. We have
\be 
\begin{aligned}
\delta \hat{h}_{\alpha\beta}=&\delta\left(\Pi_{(\alpha}^i \partial_i \partial_{\beta )} \psi + e^{-\frac{1}{2} \pb_\psi} h_{\alpha \beta} \right)\\
=&\delta\left(\Pi_{(\alpha}^i \partial_i \partial_{\beta )} \psi + \sum_{n=0}^\infty \frac{1}{(-2)^n n!} \pb^n_\psi h_{\alpha \beta} \right)\\
=&\Pi_{(\alpha}^i \partial_i \partial_{\beta )} (\lambda-\lambda^{(0)})+\pb_{(\lambda-\lambda^{(0)})} \sum_{n=1}^\infty \frac{1}{(-2)^n (n-1)!} \pb^{(n-1)}_\psi h_{\alpha \beta}\\
&\quad +\Pi_{(\alpha}^i \partial_i \partial_{\beta )} \lambda^{(0)} 
+ \sum_{n=0}^\infty \frac{1}{(-2)^n n!} \pb^n_\psi \left( - \frac{1}{2} \pb_{\lambda^{(0)}} h_{\alpha \beta} \right)\\
=&\Pi_{(\alpha}^i \partial_i \partial_{\beta )} \lambda -\frac{1}{2}\pb_{(\lambda-\lambda^{(0)})} \sum_{n=0}^\infty \frac{1}{(-2)^n n!} \pb^{n}_\psi h_{\alpha \beta}
 - \frac{1}{2} \pb_{\lambda^{(0)}}\sum_{n=0}^\infty \frac{1}{(-2)^n n!} \pb^n_\psi  h_{\alpha \beta}\\
=&\Pi_{(\alpha}^i \partial_i \partial_{\beta )} \lambda -\frac{1}{2}\pb_{\lambda} \sum_{n=0}^\infty \frac{1}{(-2)^n n!} \pb^{n}_\psi h_{\alpha \beta}\\
=& \Pi_{(\alpha}^i \partial_i \partial_{\beta )} \lambda - \frac{1}{2}  \pb_\lambda \hat{h}_{\alpha \beta} 
\end{aligned}
\ee
as needed. In the above, we repeatedly used the fact that both $\lambda$ and $\psi$ are linear in $x^i$, together with the definition of the Poisson bracket (\eqref{pb_notation} and \eqref{poisson_def}) in order to set a number of terms to $0$ as well as to commute $\pb_\psi$ and $\pb_\lambda$.

\bibliography{Ref_all_r_SD}
\bibliographystyle{utphys}

\end{document}